\documentclass[fleqn,usenatbib]{mnras}

\usepackage{newtxtext,newtxmath}
\usepackage[normalem]{ulem}

\usepackage[T1]{fontenc}

\DeclareRobustCommand{\VAN}[3]{#2}
\let\VANthebibliography\thebibliography
\def\thebibliography{\DeclareRobustCommand{\VAN}[3]{##3}\VANthebibliography}

\usepackage{graphicx}	
\usepackage{amsmath}	
\usepackage[export]{adjustbox}
\usepackage{float}

\defcitealias{Ricci_2017}{R17} 
\defcitealias{Ricci_2022}{R22}
\defcitealias{Ananna_2022a}{A22a} 
\defcitealias{Ananna_2022b}{A22b} 
\usepackage{lineno}

\title[SARD of unobscured and obscured AGN]{Accretion properties of X-ray AGN: Evidence for radiation-regulated obscuration with redshift-dependent host galaxy contribution}

\author[B. Laloux et al.]{
Brivael Laloux$^{1,2}$\thanks{E-mail: brivael.laloux@durham.ac.uk},
Antonis Georgakakis$^{2}$,
David M. Alexander$^{1}$,
Johannes Buchner$^{3,4}$,
Carolina Andonie$^{1}$,\newauthor
Nischal Acharya$^{5}$,
James Aird$^{6,7}$,
Alba V. Alonso-Tetilla$^8$,
Angela Bongiorno$^9$,
Ryan C. Hickox$^{1,10}$,
Andrea Lapi$^{11}$,\newauthor
Blessing Musiimenta$^{12, 13}$,
Cristina Ramos Almeida$^{14,15}$,
Carolin Villforth$^{16}$,
Francesco Shankar$^{8}$
\\
$^{1}$ Centre for Extragalactic Astronomy, Department of Physics, Durham University, UK\\
$^{2}$ Institute for Astronomy \& Astrophysics, National Observatory of Athens, V. Paulou \& I. Metaxa, 11532, Greece\\
$^{3}$ Max Planck Institute for Extraterrestrial Physics, Giessenbachstrasse, 85741 Garching, Germany\\
$^{4}$ Excellence Cluster Universe, Boltzmannstr. 2, D-85748, Garching, Germany\\
$^{5}$ Donostia International Physics Center (DIPC), Manuel Lardizabal Ibilbidea, 4, San Sebastián, Spain\\
$^{6}$ Institute for astronomy, University of Edinburgh, Royal observatory, Edinburgh, EH9 3HJ, UK\\
$^{7}$ School of Physics \& Astronomy, University of Leicester, University road, Leicester, LE1 7RJ, UK\\
$^{8}$ School of Physics and Astronomy, University of Southampton, Highfield, SO17 1BJ, Southampton, UK\\
$^{9}$ INAF-Observatory of Rome, via di Frascati 33, 00074, Monteporzio Catone, Rome, Italy\\
$^{10}$ Department of Physics and Astronomy, Dartmouth College, Hanover, NH 03755, USA\\
$^{11}$ SISSA, Via Bonomea 265, 34136 Trieste, Italy\\
$^{12}$ Dipartimento di Fisica e Astronomia ’Augusto Righi’, Alma Mater Studiorum - Università di Bologna, Via Gobetti 93/2, 40129 Bologna, Italy\\
$^{13}$  INAF-Osservatorio di Astrofisica e Scienza dello Spazio via Gobetti 93/3, 40129 Bologna, Italy\\
$^{14}$ Instituto de Astrofísica de Canarias, Calle Vía Láctea, s/n, E-38205 La Laguna, Tenerife, Spain\\
$^{15}$  Departamento de Astrofísica, Universidad de La Laguna, E-38206 La Laguna, Tenerife, Spain\\
$^{16}$ University of Bath, Department of Physics, Claverton Down, Bath, BA2 7AY, UK\\
}

\date{Accepted XXX. Received YYY; in original form ZZZ}

\pubyear{2015}

\begin{document}
\label{firstpage}
\pagerange{\pageref{firstpage}--\pageref{lastpage}}
\maketitle


\begin{abstract}

{
We adopt a Bayesian X-ray spectral approach to investigate the accretion properties of unobscured ($20<\log(N_{\rm H}/{\rm cm}^{-2}<22$) and obscured ($22< \log(N_{\rm H}/{\rm cm}^{-2}<24$) active galactic nuclei (AGN) to shed light on the orientation vs evolution scenarios for the origin of the obscuring material. 
For a sample of 3882 X-ray-selected AGN from the {\it Chandra} COSMOS Legacy, AEGIS and CDFS extragalactic surveys, we constrain their stellar masses, $M_\star$, intrinsic X-ray luminosities, $L_{\rm X}$, obscuring column densities, $N_{\rm H}$, and specific accretion rates $\lambda\propto L_{\rm X}/M_\star$.
By combining these observables within a Bayesian non-parametric approach, we infer, for the first time, the specific accretion rate distribution (SARD) of obscured and unobscured AGN to $z\approx3$, i.e. the probability of a galaxy with mass $M_\star$ at redshift $z$ hosting an AGN with column density $N_{\rm H}$ and specific accretion rate $\lambda$. 
Our findings indicate that 
(1) both obscured and unobscured SARDs share similar shapes, shifting towards higher accretion rates with redshift, 
(2) unobscured SARDs exhibit a systematic offset towards higher $\lambda$ compared to obscured SARD for all redshift intervals, 
(3) the obscured AGN fraction declines sharply at $\log\lambda_{\rm break} \sim-2$ for $z <0.5$, but shifts to higher $\lambda$ values with increasing redshift, 
(4) the incidence of AGN within the theoretically unstable blow-out region of the $\lambda-N_{\rm H}$ plane increases with redshift. 
These observations provide compelling evidence for AGN "downsizing" and radiation-regulated nuclear-scale obscuration with an increasing host galaxy contribution towards higher redshifts.

}

\end{abstract}




\section{Introduction}\label{sec:Introduction}

It is currently thought that most massive galaxies in the local Universe host black holes in their nuclear regions with masses that approach or may even exceed $10^{10} \, M_\odot$ \citep[e.g.][]{Kormendy_2013}. These extreme compact objects are believed to grow either through merging with other black holes \citep[e.g.][]{Volonteri_2003, ONeill_2022} or via accretion of matter from their immediate surroundings \citep[e.g.][]{Soltan_1982, Alexander_2012}. The former mechanism may dominate the low-frequency gravitational wave background signal observed by current pulsar timing arrays \citep{Antoniadis_2023} and will be studied in detail by the future LISA \citep[Laser Interferometer Space Antenna,][]{Wyithe_2003, Arun_2022} gravitational wave observatory. Although merging black holes are interesting, observational evidence demonstrates that the dominant channel for the growth of supermassive black holes (SMBHs) at the centres of galaxies is accretion \citep[e.g.][]{Soltan_1982}. During this process, gaseous material is funnelled to the close vicinity of the compact object and is accreted onto it. Simultaneously, large amounts of energy are produced and can be observed as radiation across the electromagnetic spectrum. Although SMBHs experiencing such accretion events (Active Galactic Nuclei, AGN) are being observed throughout the Universe, the details of the physical conditions that initiate them, as well as the impact of the produced energy on the host galaxy, are still debated. 

One approach to address these issues is to study, in a statistical manner, the accretion properties of AGN in relation to the characteristics of their host galaxies. Any underlying trends or covariances that such a methodology may reveal provide hints about the physical processes governing the feeding and feedback cycle of SMBHs \citep[e.g.][]{Alexander_2012}. Of particular importance in this respect is the Eddington ratio,  $\lambda_{\rm Edd}=L_{\rm bol}/L_{\rm Edd} \propto L_{\rm bol}/M_{\rm BH}$, of an accreting system. This fundamental characteristic of AGN measures how fast or slow a SMBH accretes material relative to its maximum capacity, i.e. the Eddington limit, where radiation pressure becomes dominant and may regulate the flow of matter \citep[e.g.][]{Fabian_2006, Fabian_2008, Di_matteo_2005, Fabian_2012, Ricci_2017}. Therefore, the Eddington ratio contains information on both the efficiency of the accretion process and the potential of the system to launch winds or outflows into the interstellar medium of its host galaxy. In recent years it has become possible to measure Eddington ratios or proxies of that quantity for large samples of AGN both in the local Universe \citep[e.g.][]{Kauffmann_Heckman_2009, Schawinski_2010, Ricci_2017, Ananna_2022a, Birchall_2023, Torbaniuk_2024} and at higher redshift \citep[e.g.][]{Schulze_2010, Bongiorno_2012, Nobuta_2012, Georgakakis_2014, Schulze_2015, Bongiorno_2016, Aird_2018}. A broad picture emerging from these studies is a preference for AGN residing in star-forming hosts \citep{Mullaney_2015, Scholtz_2018, Ni_2021}, possibly indicating the availability of gas reservoirs and a sharp decline in the incidence of active SMBHs close to or above the Eddington limit, suggesting self-regulation. 

Another fundamental property of AGN is the level of obscuration along the line of sight to the observer. It is well established that most SMBHs accrete material behind dust and gas clouds \citep[e.g.][]{Ueda_2014, Aird_2015, Hickox_2018, Andonie_2022} that absorb part of the emitted radiation and block the direct view to the central engine, particularly in the UV/optical part of the electromagnetic spectrum. The nature and origin of the obscuring material are not yet fully understood \cite[e.g.][]{Netzer_2015}, although it is proposed to be associated with the feeding and feedback cycle of AGN \citep[e.g.][]{Sanders_1988, Wada_2012, Wada_2015}, thereby, provide information on the physics of black hole growth. Interestingly, in this respect, recent studies on the statistical properties of AGN samples find covariances between AGN obscuration and Eddington ratio. For example, \citet[][\citetalias{Ricci_2017} hereafter]{Ricci_2017} and  \citet[][\citetalias{Ricci_2022} hereafter]{Ricci_2022}, use a hard-X-ray-selected AGN sample ($>10$\,keV) at low redshift \citep[$\bar{z}=0.037$, ][]{Koss_2022a} to show that the fraction of obscured AGN drops sharply with increasing  Eddington ratio above $\lambda_{\rm break} \approx 10^{-2}$. This trend is consistent with the expectations of AGN radiation-driven outflows acting on a dusty medium \citep{Fabian_2008,  Fabian_2009}, thereby pushing away the obscuring clouds and perhaps regulating the accretion flow onto the SMBH. 

In this paper, we expand for the first time the results above to higher redshift to explore if the trends between obscuration and Eddington ratio established in the local Universe persist to earlier cosmic times during the peak of the accretion history of the Universe \citep[redshfits $z\approx1-3$,][]{Ueda_2014, Aird_2015}.
We adopt the specific accretion-rate \citep[SAR,] []{Bongiorno_2012, Aird_2012}, i.e. the ratio between AGN X-ray luminosity and stellar mass of the host galaxy, as a proxy of the Eddington ratio. This is because the direct determination of black hole masses (and hence Eddington ratios) for obscured AGN outside the local Universe (e.g. via measurements of the velocity dispersion of the host galaxy bulge), is challenging and expensive in observing resources. Instead, galaxy stellar masses, which are believed to correlate with the masses of the black holes at their nuclear regions \citep{Marconi_2003, Kormendy_2013, Suh_2020}, can be measured for large samples of AGN hosts via template fits to their observed Spectral Energy Distribution (SED). Although this is a clear advantage, one should also bear in mind that the scaling from galaxy stellar mass to black hole mass includes assumptions and potential biases \citep{Lopez_2023}. Although extensive work has been carried out on the stellar mass vs black hole mass relation in the local Universe \citep[e.g.][]{Kormendy_2013}, the redshift evolution of this scaling law and its scatter remains poorly constrained. The impact of these uncertainties in the analysis is discussed, when relevant, in our paper.

We also use X-ray surveys to define the parent AGN sample used in the analysis. This is because X-ray observations (i) provide a handle on the line-of-sight obscuration to the active SMBHs (parametrised by the atomic hydrogen column density $N_{\rm H}$) and (ii) have a selection function that can be accurately quantified, thereby allowing inference on the demographics of the underlying population from an observed X-ray AGN sample \cite{Brandt_2015}. Our analysis also measures the incidence of AGN in galaxies as a function of obscuration. For this exercise, we infer from the observations the specific accretion rate distribution (SARD) of AGN, which measures the probability of a galaxy hosting an active black hole of a given SAR. This analysis step extends previous works on the overall SARD of the AGN population \citep[][]{Aird_2012, Bongiorno_2012, Georgakakis_2017b, Aird_2018} that do not discriminate between systems with different levels of line-of-sight obscuration. Section\,\ref{Sec:Data_reduc} describes the X-ray surveys and the multi-wavelength data used in this work. Section\,\ref{sec:Analysis} discusses the analyses of the observations to derive line-of-sight obscuration levels and galaxy stellar masses and infer the AGN incidences. We present our results in Section\,\ref{Sec:Results} and show how the SARD evolves with redshift and obscuration. In Section \ref{Sec:Discussion}, we discuss the implications of our results on the fueling mechanisms of SMBHs and their evolution. Finally, our findings are summarized in Section\,\ref{Sec:Summary}. Throughout the paper, we adopt the following cosmological parameters: $H_0 = 70$\,km\,s$^{-1}$\,Mpc$^{-1}$, $\Omega_{\rm M}=0.3$ and $\Omega_{\rm \Lambda}=0.7$.

\section{Data}\label{Sec:Data_reduc}

\subsection{Multiwavelength photometry and redshift}\label{subsec:data_redshift}

We use X-ray selected AGN detected in three extragalactic survey fields observed by the \textit{Chandra} X-ray observatory \citep{Weisskopf_2000}: the \textit{Chandra} COSMOS Legacy \citep{Civano_2016}, the All-wavelength Extended Groth Strip X-ray Deep \citep[AEGIS-XD;][]{Nandra_2015} and the \textit{Chandra} Deep Field South \citep[CDFS;][]{Luo_2017}. Details regarding the construction of X-ray source catalogues, identification of the sources' multi-wavelength counterparts, and redshift determination are presented in \cite{Georgakakis_2017b}. This section provides a brief overview of the three fields and relevant data analysis steps.

The \textit{Chandra} COSMOS Legacy field is the largest survey used in our analysis with a total area of $2\,\rm{deg}^2$. The tiling of the \textit{Chandra} pointings within this field is such that the total exposure time reaches a homogeneous limit of 160\,ks in the central $\sim 1.5\,\rm{deg}^2$ part of the survey and drops to 80\,ks in the outer regions. AEGIS-XD, covering $0.29\,\rm{deg}^2$, has a longer exposure time of 800\,ks per \textit{Chandra} pointing. CDFS is the smallest of the three surveys with a $0.13\,\rm{deg}^2$ area but also the deepest with a total exposure time that sums up to a total of 7\,Ms \citep{Luo_2017}. In this work, we use the CDFS X-ray source catalogue presented by \cite{Georgakakis_2017b} that uses the first 4\,Ms of exposure time in this field (\textit{Chandra} observations carried out up to July 22, 2010) to define the X-ray selected AGN sample. For extracting X-ray spectra of the detected X-ray sources, however, all \textit{Chandra} pointings in CDFS with a total exposure time of 7,Ms are used.

The X-ray observation reduction and source detection follow the methods of \cite{Laird_2009} and \cite{Nandra_2015}. They enable accurate characterisation of the X-ray selection function, i.e. the probability of detecting an X-ray source with a given count-rate or flux across the survey area, based on the sensitivity map construction steps described in \cite{Georgakakis_2008}. X-ray sources are detected independently in four energy bands, 0.5-2\,keV (soft), 2-7\,keV (hard), 4-7\,keV (ultra-hard) and 0.5-7\,keV (full), each with the Poisson false detection probability threshold $<4\cdot10^{-6}$. Table\,\ref{tab:nb_sources_per_band} reports the total number of sources in each energy band for the three {\it Chandra} survey fields. For the purpose of optimising statistics, this study focuses on the energy band with the highest number of sources, i.e. the full band 0.5-7\,keV. Our final sample contains 3882 sources, with 2658, 806  and 418 detections within the COSMOS Legacy, AEGIS-XD and CDFS, respectively.

The identification of X-ray sources with multiwavelength counterparts in the optical to mid-infrared (mid-IR) is based on the likelihood ratio method \citep[LR;][]{Sutherland_1992, Brusa_2007}. Specific details on the identification of sources in the \textit{Chandra} survey fields used in this work can be found in \cite{Aird_2015}. The LR measures the ratio between the probability of a multiwavelength source being the true counterpart of an X-ray detection and the probability of it being a spurious association. The calculation takes into account both the magnitude of a potential counterpart and its angular separation relative to the X-ray source under consideration. The photometric catalogues used to identify X-ray sources differ for each field. In the case of the COSMOS-Legacy field, we use the COSMOS Intermediate and Broad Band Photometry Catalogue 2008\footnote{\href{http://irsa.ipac.caltech.edu/data/COSMOS/datasets.html}{http://irsa.ipac.caltech.edu/data/COSMOS/datasets.html}}. For AEGIS-XD and CDFS, counterparts are identified using a custom version of the Rainbow Cosmological Surveys Database\footnote{\href{https://arcoirix.cab.inta-csic.es/Rainbow\_Database/Home.html}{https://arcoirix.cab.inta-csic.es/Rainbow\_Database/Home.html}} \citep{Perez_Gonzalez2005, Perez_Gonzalez2008, Barro2011}, which provides a compilation of the various photometric data sets.

Spectroscopic or photometric redshifts for the X-ray sources in the \textit{Chandra} survey fields used in this work are from \cite{Hsu2014} (CDFS), \cite{Nandra_2015} (AEGIS-XD) and \cite{Georgakakis_2017b} (COSMOS Legacy).  Table\,\ref{tab:redshift} summarises the redshift information available for each field. Overall, 2122 sources have a spectroscopic redshift estimation, whereas 1721 sources have a photometric redshift probability distribution function (PDF). The remaining 39 sources do not have any redshift estimation.

\begin{figure}
    \centering
    \includegraphics[width=0.45\textwidth]{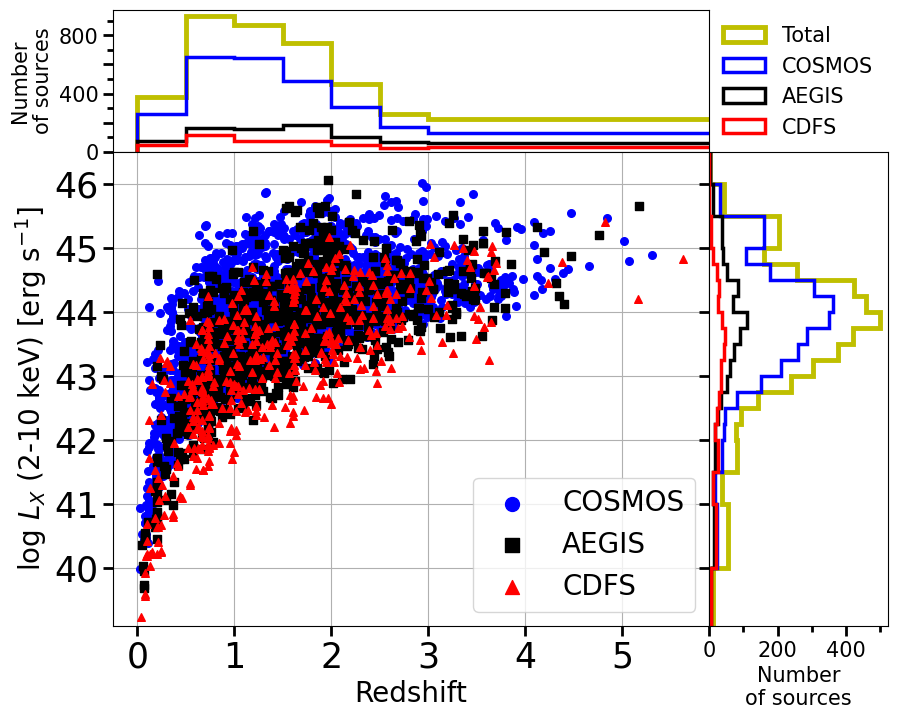}
    \caption{{\it Central panel}: Distribution of the intrinsic 2-10\,keV X-ray luminosity (see Section \ref{subsec:xray_spectroscopy}) as a function of redshift. The blue circles, black squares and red triangles correspond to 0.5-7\,keV selected X-ray sources in the COSMOS Legacy, AEGIS-XD and CDFS surveys, respectively. {\it Top panel}: Redshift distribution of the X-ray sources. The different colours are for COSMOS (blue), AEGIS-XD (black), CDFS (red) and the full sample (yellow). {\it Right panel}: Intrinsic X-ray luminosity distribution of X-ray sources in the different surveys. The colour coding is the same as in the top panel.}
    \label{fig:Lx_vs_z}
\end{figure}

Figure \ref{fig:Lx_vs_z} presents the distribution of the X-ray sources used in this work across the luminosity-redshift plane. The determination of 2-10\,keV luminosities is based on the X-ray spectral analysis results described in Section \ref{subsec:xray_spectroscopy}. Figure \ref{fig:Lx_vs_z} demonstrates that the combination of the three \textit{Chandra} survey fields with different X-ray depths and areas allows us to explore the AGN population over a reasonably broad luminosity and redshift baselines. 

\begin{table}
    \centering
    \caption{Number of unique X-ray detections in different energy bands in each of the three {\it Chandra} X-ray survey fields used in this work.}
    \begin{tabular}{c|c|c|c|c|c}
        Survey & Full & Soft & Hard & Ultra-hard & Any \\
         & 0.5-7 keV &  0.5-2 keV &  2-7 keV &  4-7 keV & band \\ 
         \hline
         \hline 
         COSMOS   & 2658 & 2172 & 1754 & 791 & 2846 \\
         \hline
         AEGIS-XD & 806  & 683  & 547  & 287 & 878 \\
         \hline
         CDFS     & 418  & 406  & 282  & 183 & 486 \\ 
         \hline
         \hline
         All      & 3882 & 3261 & 2583 & 1261 & 4210 \\         
    \end{tabular}
    \label{tab:nb_sources_per_band}
\end{table}

\begin{table}
    \centering
    \caption{Number of X-ray sources selected in the 0.5-7\,keV band with spectroscopic, photometric or no redshift estimation for each of the three {\it Chandra} X-ray survey fields used in this work.}
    \begin{tabular}{c|c|c|c|c}
         Redshift type & COSMOS & AEGIS-XD & CDFS & Total \\
         \hline
        Spectroscopic & 1536 & 319 & 267 & 2122 \\
        Photometric   & 1087 & 487 & 147 & 1721 \\
        No estimation & 35   & 0   &  4  & 39 \\ \hline
        Total         & 2658 & 806 & 418 & 3882 \\
    \end{tabular}    
    \label{tab:redshift}
\end{table}

\subsection{The X-ray spectral extraction}\label{subsec:spectral_extraction}

The extraction of X-ray spectra for individual X-ray sources follows the methodology of \cite{Laloux_2023}. The COSMOS, AEGIS-XD and CDFS surveys used in this work encompass multiple overlapping observations. Consequently, each source typically lies in several distinct {\it Chandra} pointings. Our approach extracts X-ray spectra at both the position of a given source and appropriately selected background positions from all the available overlapping observations. Calibration files, i.e. the Auxiliary Response Files (ARF) and Redistribution Matrix Files (RMF), are also produced for each of the extracted spectra of a given source. Subsequently, these products are coadded using the \texttt{combine\_spectra} task of CIAO 4.13 \citep{Fruscione_2006} to generate a single spectrum for each source, along with the associated ARF, RMF and background spectrum. The circular region within which a source's X-ray photons are extracted has a radius that maximises its signal-to-noise ratio, as explained in \cite{Laloux_2023}. The background region is an annulus surrounding the source under consideration with inner and outer radii that are respectively $2.5^{\prime\prime}$ and $20^{\prime\prime}$ larger than the extraction radius of the source's X-ray spectrum. The outer radius may be progressively expanded until the background region contains at least 100 photon counts. Any other X-ray sources that lie within the background region are masked out during this process.

\section{Analysis}\label{sec:Analysis}

This section outlines the methodological steps that lead to the determination of the SARD for obscured and unobscured AGN from the X-ray selected sample of Section\,\ref{Sec:Data_reduc}. The Bayesian X-ray spectral fitting, described in Section\,\ref{subsec:xray_spectroscopy}, yields PDF for the intrinsic 2-10\,keV X-ray luminosity, $L_{\rm X}$, and line-of-sight column density, $N_{\rm H}$, for individual X-ray AGN. The stellar masses, $M_\star$, of the AGN host galaxies are estimated by fitting templates to their broadband SED as described in Section\,\ref{subsec:SED_fitting}. The statistical Bayesian inference approach, combining the outcomes of the X-ray spectral analysis and SED fitting process to derive the SARD as a function of the obscuration, is described in Section\,\ref{subsec:sBHAR_calculation}.

\subsection{X-ray spectral fitting}\label{subsec:xray_spectroscopy}

In this work, we use the Bayesian X-ray Analysis (BXA) package \citep{Buchner_2014} to fit the X-ray spectra extracted as described in Section\,\ref{subsec:spectral_extraction}. Powered by the Ultranest package \citep{Buchner_2021}, BXA efficiently explores the entire parameter space without getting stuck in local minima. A notable advantage of this Bayesian approach is the ability to incorporate prior knowledge about specific parameters, such as the full photometric redshift PDF. 
Additionally, the output of this algorithm is in the form of a list of equiprobable sets of parameter values. From this list, one can build the posterior distribution for each free parameter of the model, accurately representing parameter uncertainties and covariances.
These posterior distributions are further propagated in the analysis, e.g. for the derivation of SARD.

Given the limited number of photons in many X-ray spectra, we adopt CSTAT statistics \citep{Cash_1979} as our likelihood function. This statistical approach accounts for the Poisson nature of the data and necessitates modelling both the source and background spectra. To address the latter, we employ the \texttt{automatic\_background()} command in BXA \citep{Simmonds_2018}. This machine-learning-based algorithm empirically derives the principal components of the background model and their parameter values.

The source spectra are fitted with the physically motivated model UXCLUMPY \citep{Buchner_2019} in which the obscuring material is distributed in clumps in a toroidal shape around the SMBH. The clumpiness of the torus is supported by observational evidence in both X-ray \citep{Risaliti_2002, Esparza_2021} and mid-infrared \citep{Ramos_2009, Ramos_2011} studies. In this model, two types of clouds surround a central emitting source, small clouds scattered in a torus-like volume and large Compton-thick (CTK, $N_{\rm H}>10^{24} {\rm \,cm}^{-2}$) clouds distributed in a ring on the equatorial plane. The UXCLUMPY spectral model is produced by a radiative transfer code that tracks individual photons emitted by a central source, interacting or not with the surrounding material before reaching the observer. This approach generates a complex spectrum dependent on the various model parameters controlling the geometry of the obscurer and the physical properties of the source. The UXCLUMPY model incorporates three main components contributing to the overall spectrum. The transmitted component represents photons that escape without interacting with the obscuring material. The reflected component includes photons that undergo scattering or reflection after interacting with the obscuring material. The fluorescent lines component represents photons that are re-emitted after absorption, with the dominant contribution from neutral iron atoms at 6.4 keV. 
Additionally, in local obscured Seyferts, a soft X-ray photon excess is often observed. This is believed to be produced by photons scattered by photo-ionized gas clouds located above the torus into line of sight of the observer \citep{Bianchi_2006}. To account for this excess, a second UXCLUMPY model is added to mirror the intrinsic spectrum without absorption. Except for its normalisation, all the parameters of this second model (e.g. spectral index) are linked to the main model.

\begin{figure*}
    \centering
    \includegraphics[width=1\linewidth]{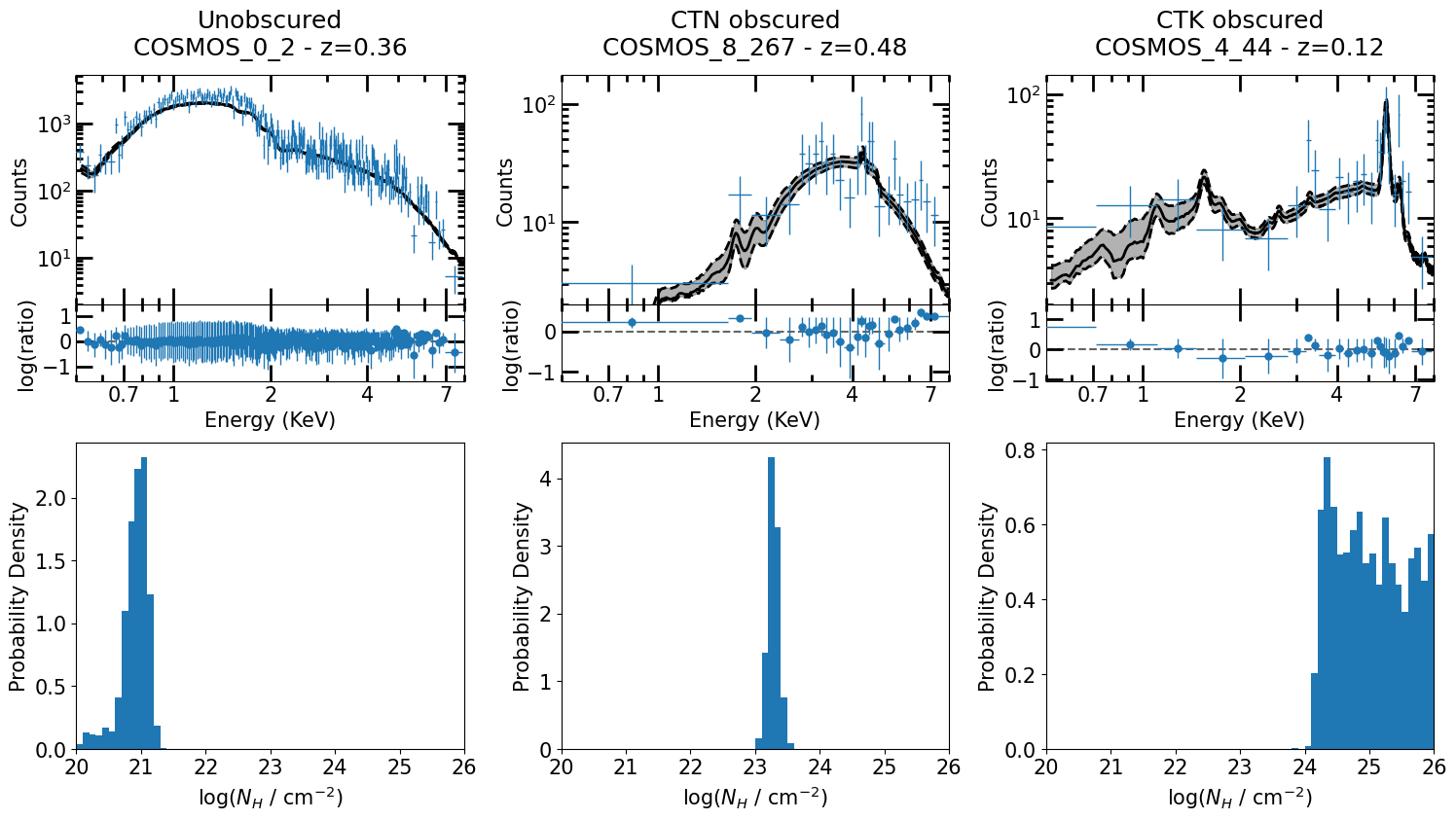}
    \caption{Examples of X-ray spectroscopic results for an unobscured, CTN obscured and CTK obscured AGN ({\it left}, {\it central} and {\it left columns}, respectively). For each source, the top panel shows the spectrum ({\it blue crosses}) and the best-fit model ({\it black solid line}), on top of the residual, defined as $\log(\frac{\rm data}{\rm model})$. The 1-$\sigma$ uncertainty of the best-fit model is shown with the {\it grey area}. The lower panels indicate for each source the $N_{\rm H}$ posterior PDF.    
    }
    \label{fig:dif_Xray_fits}
\end{figure*}

In our study, several UXCLUMPY parameters are fixed, while the others are allowed to vary within predefined prior distributions. A detailed discussion of these choices can be found in \cite{Laloux_2023}, and the parameter settings are presented in Table \ref{tab:parameters_uxclumpy}. In UXCLUMPY, the spatial distribution of the small clouds surrounding the central engine is controlled by the TORsigma parameter, which is fixed at 28°. The covering factor of the large equatorial CTK clouds is set to CTKcover\,=\,0.4. The inclination angle with respect to the plane of the torus is fixed to 45°. 
The central source of the AGN emits a power law continuum with a high energy cutoff fixed at E\_cutoff=200\,keV and a photon index, $\Gamma$, that follows a Gaussian prior distribution centred on 1.95 with a standard deviation of 0.15 \citep{Nandra_1997}. The line-of-sight column density, $N_{\rm H}$, the power-law continuum normalisation, and the soft-scattering component normalisation follow a log-uniform prior in the intervals 20 to 26, $-8$ to $3$ and $-7$ to $-1.5$, respectively. The soft-scattering normalisation refers to the fractional contribution of the soft-scattering component relative to the intrinsic power-law component. Computationally, this is achieved by setting the normalisation of the soft-scattering model to log(norm\_torus) + log(norm\_scattering). 
If a source has a spectroscopic redshift, the UXCLUMPY redshift parameter is fixed to that value. If a photometric redshift PDF is available, it is used as a prior in the modelling. If no redshift information is available (a total of 39 sources, see section\,\ref{subsec:data_redshift}), then we assign a uniform prior in the range $1<z<6$ to the UXCLUMPY redshift parameter as a source for which we cannot obtain a redshift estimation is most likely above $z>1$ \citep{Georgakakis_2017b}. The normalization of the background model is set to the value estimated by the \texttt{automatic\_background()} task of BXA re-scaled to the source area. This decision does not impact the final results. 
The X-ray spectroscopic methodology described above, by accounting for observation uncertainty, yields robust parameter constraints \citep{Laloux_2023}. 
Figure\,\ref{fig:dif_Xray_fits} illustrates the spectroscopic results by displaying the best-fit model of the spectra of three sources (from left to right, an unobscured, a CTN obscured and a CTK obscured AGN).

\begin{table}
    \centering
    \caption{Input parameters of the UXCLUMPY model and the adopted fixed value or the prior used in the X-ray spectral analysis.}
    \begin{tabular}{l|l|}
        parameter & prior \\ \hline
        TORsigma & fixed =  28°\\ 
        CTKcover & fixed = 0.4\\ 
        Inclination  & fixed = 45°\\
        E\_cutoff & fixed = 200\,keV\\ 
        photon index $\Gamma$ & Gaussian(1.95, 0.15)\\ 
        ${\rm log}(N_{\rm H}/{\rm cm}^{-2})$   & uniform(20, 26)\\ 
        log(norm\_torus) & uniform(-8, 3) \\
        log(norm\_scattering)  & uniform(-7, -1.5)\\
        redshift spectroscopic & fixed at source redshift \\ 
        redshift photometric & photometric PDF \\
        no redshift estimation & uniform(1, 6) \\
        log(norm\_background)  & fixed at measured value
        
    \end{tabular}

    \label{tab:parameters_uxclumpy}
\end{table}

\subsection{Spectral energy distribution fitting}\label{subsec:SED_fitting}

Fitting theoretical and/or empirical model templates to the observed SED of extragalactic sources yields valuable information on their physical properties \citep{Conroy_2013}. In this study,  our primary goal is to recover the stellar masses of AGN host galaxies by modelling their multiwavelength (UV to mid-IR) broad-band photometry. The templates we adopt include both stellar population and AGN emission modules, enabling the separation of galaxy light from that associated with the nuclear compact object and thus allowing the estimation of critical galaxy properties, such as stellar mass. For this application, we choose to use the Code Investigating GALaxy Emission \citep[CIGALE hereafter, ][]{Boquien_2019, Yang_2022} because it is computationally efficient and extensively tested.

CIGALE provides different SED template-generating modules, each corresponding to different emission mechanisms and/or components. These modules are governed by a set of parameters that can vary within a user-defined grid. The combination of the different modules results in a multi-dimensional parameter grid where each node represents a specific combination of parameter values for which CIGALE generates a template SED. The comparison of these models with the observed photometry using the $\chi^2$ statistic yields the best-fit set of module parameters, their associated uncertainties and their PDFs. To ensure the selection of appropriate modules and parameter sets, we follow the prescriptions of previous studies \citep[][see Table \ref{tab:cigale_params} in the Appendix]{Mountrichas_2021, Lopez_2023}. The chosen modules and their corresponding parameters are briefly described below.

For the host galaxy stellar component, we employed a delayed star-formation history that parameterises the star-formation rate (SFR) as ${\rm SFR}(t) \propto t\times{\rm exp}(-t / \tau)$. The SFR peaks at the time $\tau$ and then declines exponentially, with an optional star formation burst with an e-folding time of 50 Myr to account for recent intense star-formation activity. 
The stellar population is modelled from the stellar templates of \cite{Bruzual_2003} with the initial mass function of \cite{Chabrier_2003} and solar metallicity (Z=0.02). We also incorporate the nebular module \citep{Inoue_2011}, which includes the typical nebular emission lines observed in star-forming galaxies with a line width of 200 km s$^{-1}$. 
The dust attenuation follows a modified \cite{Calzetti_2000} law, and the dust emission from the host galaxy with no AGN contribution is modelled by the \cite{Dale_2014} templates. 
For the AGN emission, we employ the SKIRTOR model \citep{Stalevski_2016}, which uses a 3D radiation-transfer code to simulate the interaction of photons emitted by a central source with a two-phase medium torus, consisting of a smooth and low-density medium containing high-density clumps. The scaling factor of the module, the AGN fraction, represents the ratio of the IR emission from the AGN to the total emission (AGN + host galaxy) within the $8-1000 \mu {\rm m}$ rest-frame range. In this work, we consider both type I and type II AGN, corresponding to SKIRTOR model viewing angles of 30$^{\circ}$ and 70$^{\circ}$, respectively.

\begin{figure}
    \centering
    \includegraphics[width=0.9\linewidth]{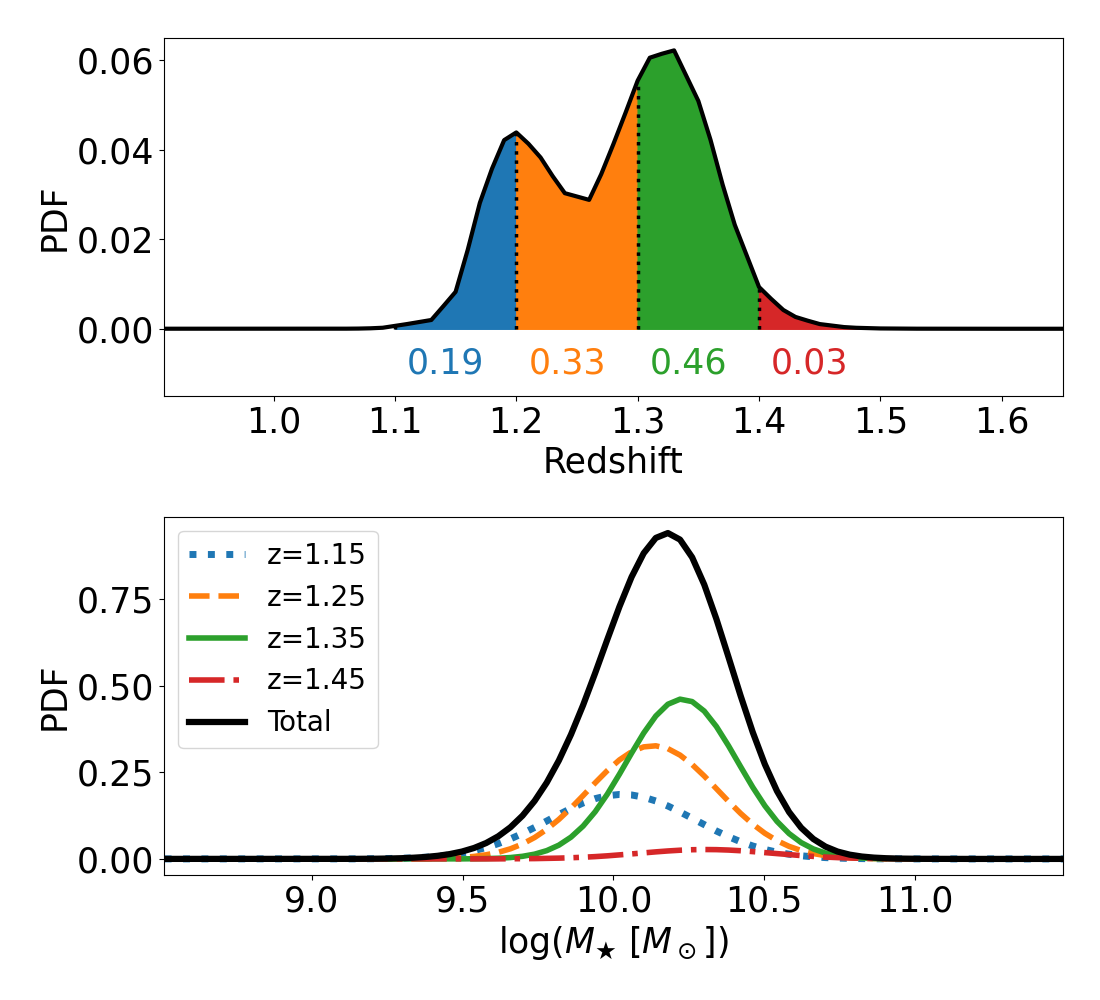}
    \caption{Demonstration of how we use CIGALE to account for photometric redshift PDFs of individual sources. 
    {\it Top panel}: The redshift PDF of an example X-ray source in the sample is split into four intervals with width $\Delta \rm z=0.1$. For clarity, the part of the PDF within each redshift interval is shaded with a different colour. The fraction of the photometric redshift PDF in each interval is indicated by coloured numbers at the bottom of the panel.
    {\it Bottom panel}: Demonstration of the Gaussian mixture model adopted to recover the stellar mass PDF for the same source. For each redshift interval of the top panel, CIGALE returns a best-fit value for the stellar mass, $M_\star$, and the corresponding uncertainty. These are used to generate Gaussians with means equal to the best fit stellar masses and standard deviations equal to the stellar mass uncertainties. These Gaussians are shown with the coloured curves. The colour coding corresponds to the redshift intervals of the top panel. The normalisation of these curves is set by the fraction of the photometric redshift PDF within each redshift interval (i.e. the coloured numbers in the top panel). The sum of the Gaussians (black solid line) is the total log($M_\star$) PDF.}
    \label{fig:zphot_SED}
\end{figure}

One of the limitations of CIGALE is that it requires fixing the redshift, preventing the utilization of photometric redshift PDFs. Given that a sizeable fraction of the X-ray selected AGN sample in this study has photometric redshift estimations, we need to address this limitation and account for the redshift uncertainties in the derivation of the physical parameters from the SED fitting process. 
Our approach consists of marginalizing over the redshift uncertainty by importance sampling. We sample the z-PDF in regular redshift intervals, and for each, we perform the SED fitting with the redshift fixed to the mean value of the interval. This approach yields, for a given source, multiple estimates of physical parameters, e.g. stellar mass or SFR, each corresponding to a different redshift bin. These estimates are then weighted and combined to generate posteriors for the parameter of interest. 

Figure\,\ref{fig:zphot_SED} illustrates our approach. We segment the redshift range into bins of 0.1 width and calculate their weight based on the fraction of the redshift PDF within each bin. For each source and each redshift bin with a weight exceeding a $10^{-3}$ threshold, we create duplicates of the photometric data, fixing the redshift to the bin's mean value, and perform SED fitting on each duplicate. Finally, for a given source, the posterior PDF of a parameter of interest, such as the stellar mass, is constructed as a Gaussian mixture. A Gaussian is generated for each redshift bin, centred on the parameter best-fit value from CIGALE. The standard deviation and the amplitude of the Gaussian are given by the uncertainty of the best-fit value and the weight of the redshift bin, respectively. The full posterior parameter distribution is the sum of these Gaussians, allowing easy extraction of the parameter's median and lower/upper limits. The stellar mass distribution of the sources of the three different surveys as a function of the redshift is illustrated in the top panel of Figure\,\ref{fig_app:Mstar_lambda_z_distrib} in the Appendix\,\ref{appendix_B}.

\subsection{Specific accretion rate distribution calculation}\label{subsec:sBHAR_calculation}

\begin{figure}
    \centering
    \includegraphics[width=0.45\textwidth]{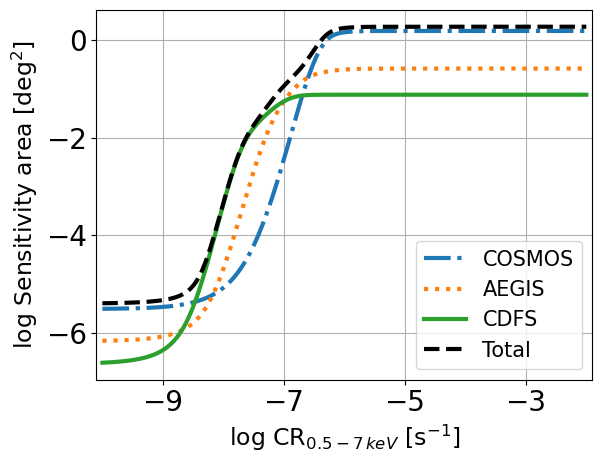}
    \caption{ Sensitivity curve as a function of count rate in the full band (0.5--7\,keV) for the COSMOS (dash-dotted blue), AEGIS-XD (dotted orange) and CDFS (solid green) surveys. The total sensitivity area for the three fields is shown with the dashed black curve.}
    \label{fig:sensi_maps}
\end{figure}

This section describes the algorithm adopted in this work to estimate the SARD of the X-ray selected AGN from Section \ref{Sec:Data_reduc} based on their inferred X-ray luminosities (refer to Section \ref{subsec:xray_spectroscopy}) and stellar masses (refer to Section \ref{subsec:SED_fitting}).
For each source, the posterior parameter distribution of both analyses must be merged to obtain the SAR $\lambda_{SAR}=L_{\rm X}/M_\star$ posterior distribution. 
For this, we attribute a stellar mass value to each parameter set $(N_{\rm H},\,L_{\rm X},\,z,\,\Gamma)$ constituting the X-ray spectroscopy posterior distribution.
This $M_\star$ value is randomly drawn from a normal distribution parameterised by the SED results. For sources with spectroscopic redshift, the centre and the width of the normal distribution correspond respectively to the best-fit stellar mass and its 1-$\sigma$ uncertainty. 
For sources with photometric redshift, the posteriors are matched by redshift, meaning the process is identical but the redshift value within the parameter sets indicates which SED fitting result to use, i.e. the one of the nearest redshift bin.
Now that each parameter set includes both the stellar mass and X-ray luminosity, the $\lambda_{SAR}$ value can be computed, resulting in the SAR posterior distribution for each individual source.
Similarly to previous works \citep{Bongiorno_2016, Georgakakis_2017b, Aird_2018}, we re-scale the $\lambda_{SAR}$ to Eddington ratio units using the bolometric correction $L_{\rm bol} = 25\cdot L_{\rm X}(\rm 2-10\,keV)$ \citep[e.g.][]{Elvis_1994} and the scaling relation $M_{\rm BH} = 0.002 M_\star$ \citep{Marconi_2003} in the following way 

\begin{align}\label{eq:sar2edd}
    \begin{split}
        \lambda &= \frac{L_{\rm bol}}{L_{\rm Edd}} = \frac{L_{\rm bol}}{1.3 \times 10^{38} \cdot M_{\rm BH}} \approx \frac{25 \cdot L_{\rm X} {\rm (2-10 keV)}}{1.3 \times 10^{38} \cdot 0.002 M_\star } \\
    &\sim 10^{-34} \,\frac{L_{\rm X}}{M_\star},
    \end{split}
\end{align}

\noindent with $M_\star$ in solar mass units and X-ray luminosity in ${\rm erg}\,{\rm s}^{-1}$. We emphasise that these simplistic scaling relations are applied to make the accretion rate $L_{\rm X}/M_\star$ resemble the Eddington ratio. The impact of these assumptions on our results and conclusions is discussed in later sections. Figure\,\ref{fig_app:Mstar_lambda_z_distrib} in Appendix\,\ref{appendix_B} illustrates the distribution of individual X-ray sources in the COSMOS, AEGIS-XD and CDFS surveys on the $\lambda$ vs redshift parameter space.

The estimation of the SARD relies on the sampling of the Poisson likelihood function originally used to determine the X-ray luminosity function of AGN \citep[XLF;][]{Loredo_2004, Aird_2010, Aird_2015, Buchner_2015, Georgakakis_2015, Laloux_2023}. We exploit the fact that the  AGN XLF can be expressed as the convolution of the stellar mass function (SMF), $\psi(M_\star| z)$, and the SARD, $P(\lambda | z, M_\star)$, which is defined
as the probability of a galaxy at redshift $z$ and with stellar mass
$M_\star$ hosting an AGN accreting at a specific accretion rate $\lambda$ \citep{Georgakakis_2017a}.
We extend this framework to include an additional dependence of the SARD on AGN obscuration, which is parameterised by the line-of-sight hydrogen column density $N_{\rm H}$. With this extra dimension, the likelihood function can be written as

\begin{align}
    \begin{split}
        \mathcal{L}(\mathcal{D}\,|\,\omega) = e&^{-\mu(\omega)} \times 
        \prod_{i=1}^N  \int \frac{dV}{dz}dz\; d{\rm log}M_\star \; d {\rm log} N_{\rm H}\; d {\rm log} \lambda \\
        & p(z, M_\star, N_{\rm H}, \lambda\, |\, d_i)\; \psi(M_\star\, | \,z)\; P(\lambda, N_{\rm H}, z, M_\star \, |\, \omega) ,
    \end{split}
    \label{eq:Likelihood}
\end{align}

\noindent where $p(z, M_\star, N_{\rm H}, \lambda\,|\, d_i)$ is the probability that a source has a redshift $z$, a mass $M_\star$, an obscuration $N_{\rm H}$ and an accretion rate $\lambda$ given the observational data $d_i$. $P(\lambda, N_{\rm H}, z, M_\star \, |\, \omega)$ is the model of the SARD (at fixed $\lambda$, $N_{\rm H}$, $z$, $M_\star$) that is to be constrained by the observations and is described by the set of parameters $\omega$. The SARD is a probability density function and, therefore, at fixed redshift and $M_\star$ interval, it follows the normalisation

\begin{equation}\label{eq:normalisation}
\int  P(\lambda, N_{\rm H}, z, M_\star) \, d\log\lambda \;d\log N_{\rm H} =1.
\end{equation}

\noindent The expected total number of detected AGN in a survey for a particular set of model parameters $\omega$ is 

\begin{align}
    \begin{split}
        \mu(\omega) = \int &\frac{dV}{dz}dz\; d{\rm log}M_\star \; d {\rm log} N_{\rm H}\; d {\rm log} \lambda \\
             & A(\lambda, N_{\rm H}, z, M_\star)\; \psi(M_\star\,|\,z)\; P(\lambda, N_{\rm H}, z, M_\star \, |\, \omega),
    \end{split}
    \label{eq:mu}
\end{align}

\noindent where $A(\lambda, N_{\rm H}, z, M_\star)$ is the sensitivity curve of the X-ray selected sample, quantifying the probability of detection of a source with a given $N_{\rm H}$, $\lambda$, $z$ and $M_\star$. It is derived from the individual sensitivity maps generated for each survey, following \cite{Georgakakis_2008}. These maps represent the area sensitive to a source with a given 0.5--7\,keV X-ray photon count rate, and their sum gives the total sensitivity map, as displayed in Figure\,\ref{fig:sensi_maps}.
Then, following \cite{Laloux_2023}, we predict with the UXCLUMPY spectral model the expected count rate for a source for a given parameter set $(\lambda, N_{\rm H}, z, M_\star)$. 
Combining the expected count rates with the total sensitivity map, normalised to the total survey area, yields the sensitivity curve $A(\lambda, N_{\rm H}, z, M_\star)$.
In the equations above, we adopt the stellar mass function parametrisation presented by \cite{Ilbert_2013}, which is consistent with recent observational constraints \citep[e.g.][]{Weaver_2023}. \cite{Ilbert_2013} use two Schechter functions with parameters evolving with redshift to represent the mass function of galaxies in the redshift interval $z = 0-4$.

For computational efficiency, we adopt a non-parametric approach for the determination of $P(\lambda, N_{\rm H}, z, M_\star)$. This involves defining a 4-dimensional grid in SAR, column density, redshift and stellar mass, where the SARD is assumed to be constant within each grid hypercube with dimensions ($\log \lambda \pm \delta\log \lambda$, $N_{\rm H}\pm \delta N_{\rm H}$, $z\pm \delta z$, $M_\star\pm \delta M_\star$). To ensure smoothness, a prior is imposed on the SARD slope along the $\lambda$-axis. 
This prior enforces that the inclination of the slope between two successive grid points is drawn from a Gaussian distribution centred on the slope inclination between the two preceding grid points.

Table\,\ref{tab:space_parameter_bins} lists the grid boundaries of the hypercubes that define the non-parametric $P(\lambda, N_{\rm H}, z, M_\star)$ model. Although included for consistency, the highest redshift bin ($3<z<6$) is not utilized in our analysis due to its limited number of sources. The logarithmic column density space is split into three bins: unobscured ($20< \log N_{\rm H}/{\rm cm}^{-2} < 22$), obscured ($22< \log N_{\rm H}/{\rm cm}^{-2} < 24$) and Compton-thick (CTK, $24< \log N_{\rm H}/{\rm cm}^{-2} < 26$). However, in this paper, we focus solely on Compton Thin AGN (CTN, $\log N_{\rm H}/{\rm cm}^{-2}<24$) due to challenges in robustly identifying CTK AGN in deep extragalactic survey fields \citep[e.g.][]{Ananna_2019, Laloux_2023}. Nonetheless, we include a CTK interval in the analysis to account for the $N_{\rm H}$-PDF of individual sources extending into this regime. Along the $\lambda$-axis, the grid is finer for moderate to high accretion rates ($-2<{\rm log} \lambda<1.5$) to allow more flexibility in this regime where the SARD  may vary rapidly. The purpose of the lowest specific accretion rate bin ($-10<{\rm log} \lambda<-5$) is to account for the majority of galaxies that host practically inactive SMBHs at their nuclear regions. We adopt a single logarithmic stellar mass bin in the interval 9.5--11.5. The chosen stellar mass range brackets the majority of our sources, and a finer binning in this axis would result in more noisy SARD constraints.

\begin{table}
    \centering
    \caption{Initial boundaries of the parameter space bins.}
    \begin{tabular}{c | c}
        Parameter & Bin boundaries\\ \hline
        redshift & [0.0, 0.5, 1.0, 1.5, 2.0, 2.5, 3.0, 6.0]\\
        $\log(N_{\rm H}/{\rm cm^{-2}})$  & [20, 22, 24, 26]\\
        log$\lambda$ & [-5, -4, -3.5, -3, -2.5, -2, -1.75, -
1.5, -1.25, -1, \\& -0.75,-0.5, -0.25, 0, 0.25, 0.5, 0.75, 1, 1.25, 1.5, 2]\\
        $\log(M_\star /M_\odot)$ & [9.5, 11.5] \\
    \end{tabular}
    \label{tab:space_parameter_bins}
\end{table}

\begin{figure*}
    \centering
    \includegraphics[width=0.95\textwidth]{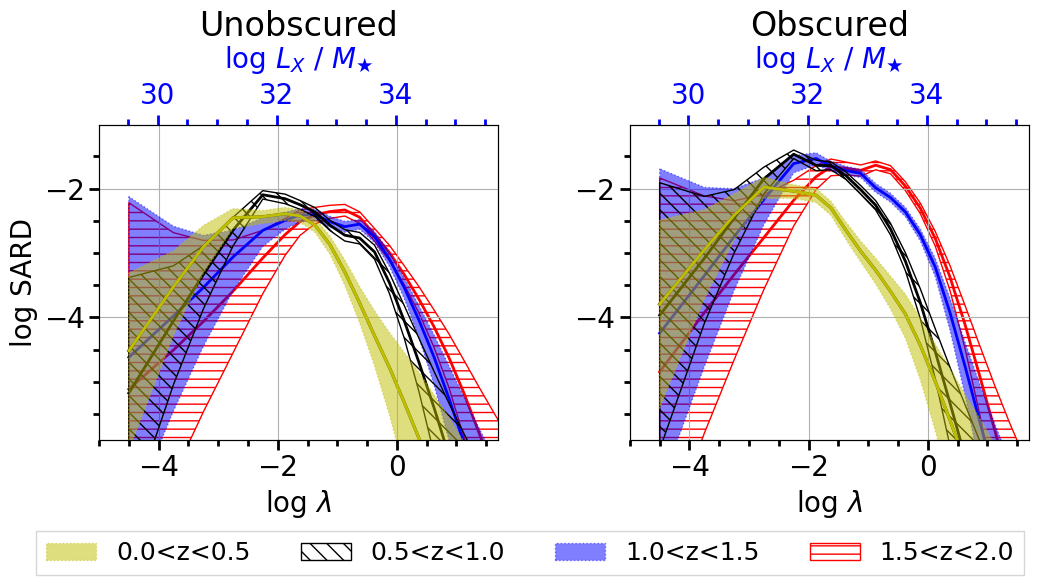}
    \caption{Specific accretion rate distribution for the unobscured (\textit{left panel}, $\log N_{\rm H}/{\rm cm}^{-2} < 22$) and obscured CTN (\textit{right panel}, $22<\log N_{\rm H}/{\rm cm}^{-2}<24$) populations. The solid lines and corresponding regions represent the median and the 1$\sigma$ confidence intervals, respectively, for different redshift bins: 0--0.5 (yellow shaded), 0.5--1 (black hatched), 1--1.5 (blue shaded) and 1.5--2 (horizontal red hatched).  The lower x-axis in each panel is the SAR rescaled to units of Eddington ratio via Equation \ref{eq:sar2edd}. For completeness, we also show the SAR in units of $\rm erg\,s^{-1}\,M_\odot^{-1}$ at the upper x-axis of each panel. }
    \label{fig:sBHAR_z_evol}
\end{figure*}

Normal galaxies with X-ray emission dominated by binaries or hot gas instead of accretion onto SMBHs are present in our sample and are likely to be relevant toward low luminosities ($L_{\rm X}<10^{42}$\,erg\,s$^{-1}$). We account for this potential source of contamination in a statistical way by adding terms in the likelihood (Equations \ref{eq:Likelihood}, \ref{eq:mu}) that depend on the X-ray luminosity function of normal (i.e. non-AGN) galaxies. The modified likelihood can be written as

\begin{align}
    \begin{split}
        \mathcal{L}(\mathcal{D}\,|\,\omega, \theta_{\rm gal}) =& e^{-\mu(\omega)} \times
        \prod_{i=1}^N \left[ \int \frac{dV}{dz}dz\; d{\rm log}M_\star \; d {\rm log} N_{\rm H}\; d {\rm log} \lambda \right.\\
        & p(z, M_\star, N_{\rm H}, \lambda\,|\, d_i)\; \psi( M_\star\,|\,z)\; P(\lambda, N_{\rm H}, z, M_\star \, |\, \omega) \\
        & + \int d{\rm log}L_{\rm X} \; \frac{dV}{dz}dz  \left. p(L_{\rm X}, z\,|\,d_i)\; \phi_{\rm gal}(L_{\rm X}, z \,|\, \theta_{\rm gal}) \right].
    \end{split}
    \label{eq:Likelihood_w_gal}
\end{align}

\begin{align}
    \begin{split}
        \mu(\omega) =& \int \frac{dV}{dz}dz\; d{\rm log}M_\star \; d {\rm log} N_{\rm H}\; d {\rm log} \lambda \\
             & A(\lambda, N_{\rm H}, z)\; \psi(M_\star|z)\; P(\lambda, N_{\rm H}, z, M_\star\,|\, \omega) \\
             & + \int d{\rm log}L_{\rm X} \; \frac{dV}{dz}dz \; A_{\rm gal}(L_{\rm X}, z) \; \phi_{\rm gal}(L_{\rm X}, z\, | \,\theta_{\rm gal}),
    \end{split}
    \label{eq:mu_w_gal}
\end{align}

\noindent where $p(L_{\rm X}, z\,|\,d_i)$ is the parameter posterior distribution from X-ray spectroscopy and $\phi_{\rm gal}(L_{\rm X}, z\, | \,\theta_{\rm gal})$ is the normal galaxy XLF, characterised by the set of parameters $\theta_{\rm gal}$. Given that the constraint of this non-AGN galaxy component is outside the scope of this work and well-studied in the past \citep[e.g.][]{Georgantopoulos_2005, Ptak_2007}, we decide not to fit this function. We use instead the parametric X-ray luminosity function $\phi_{\rm gal}(L_{\rm X}, z | \theta_{\rm gal})$ of \cite{Aird_2015}, based on the previous works of \cite{Georgakakis_2006, Georgakakis_2007}. This function has a Schechter form

\begin{equation}
    \phi_{\rm gal}(L_{\rm X}, z) = K \left( \frac{L_{\rm X}}{L_{\rm X}^*} \right)^{-\alpha} \, e^{-\frac{L_{\rm X}}{L_{\rm X}^*}},
\end{equation}

\noindent where $L_{\rm X}$ is the 2-10\,keV X-ray luminosity and $L_{\rm X}^*$ is defined as

\begin{equation}
    {\rm log}\,L_{\rm X}^*(z) = 
    \begin{cases}
     {\rm log}\, L_0 + \beta \,{\rm log}(1+z) & {\rm if}\, z<z_c \\
     {\rm log}\, L_0 + \beta \,{\rm log}(1+z) & {\rm if}\, z\geq z_c
    \end{cases}
\end{equation}
The different parameters are fixed to the best-fit values presented by \cite{Aird_2015} and listed in Table \ref{tab:param_phi_gal}.

\begin{table}
    \centering
    \caption{Parameters of the normal galaxy X-ray luminosity function $\phi_{\rm gal}$.}
    \begin{tabular}{c|c|c|c|c|c}
        Parameter & log $K$& log $L_0$ & $\alpha$& $\beta$&  $z_c$\\  \hline
        Value & -3.59& 41.12& 0.81 & 2.66& 0.82\\

    \end{tabular}
    \label{tab:param_phi_gal}
\end{table}

The Hamiltonian Markov Chain Monte Carlo code STAN \citep{STAN} is used for Bayesian statistical inference. It is used to sample the likelihood of Equation\, \ref{eq:Likelihood_w_gal} and produce posterior distributions for each of the grid hypercubes that define the non-parametric SARD model, $P(\lambda,\,N_{\rm H}\,,z,\,M_\star\, |\,\omega)$.

\section{Results}\label{Sec:Results}
\subsection{Specific accretion rate distribution}\label{subsec:sBHAR}

\begin{figure*}
    \centering
    \includegraphics[width=\textwidth]{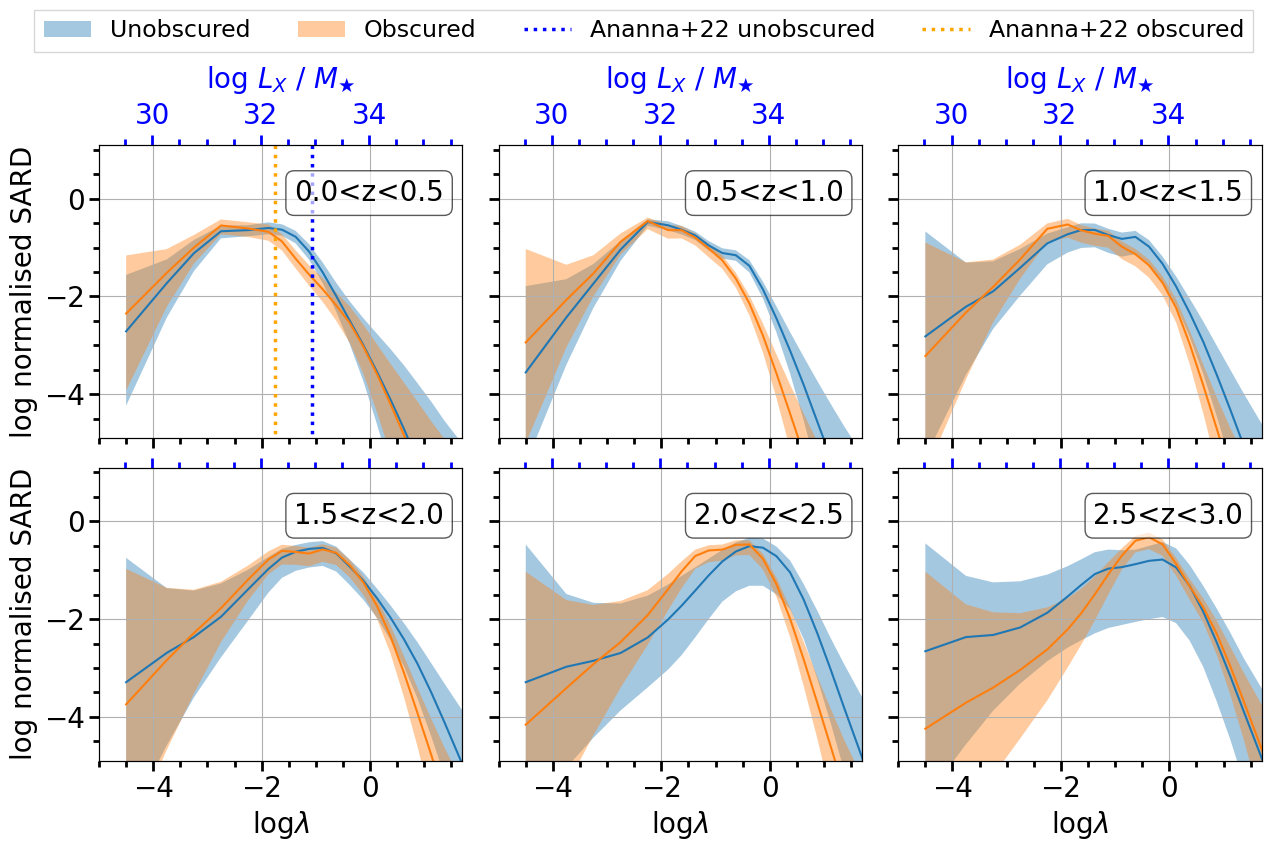}
    \caption{Specific accretion rate distributions for the unobscured ({\it blue}, $\log N_{\rm H}/{\rm cm}^{-2}<22$) and obscured ({\it orange}, $22<\log N_{\rm H}/{\rm cm}^{-2}<24$) AGN populations. The plotted SARDs are normalised to unity for each obscuration bin following Equation\,\ref{eq:normalisation-1}. The lower x-axis in each panel is the SAR rescaled to units of Eddington ratio via Equation\,\ref{eq:sar2edd}. For completeness, we also show the specific accretion rate in units of $\rm erg\,s^{-1}\,M_\odot^{-1}$ on the upper x-axis of each panel. The solid lines and shaded regions represent the median and the 1$\sigma$ confidence intervals, respectively. Each panel corresponds to a different redshift interval from $z=0-0.5$ to $z=2.5-3$. In the $z<0.5$ redshift interval, the blue and orange dotted vertical lines correspond, respectively, to the break of the Eddington ratio distribution function of unobscured and obscured AGN in \protect\citetalias{Ananna_2022b}.}
    \label{fig:sBHAR_normedNH}
\end{figure*}

Figure\,\ref{fig:sBHAR_z_evol} shows the SARDs for unobscured ($10^{20} < N_{\rm H} < 10^{22}\, \rm cm^{-2}$) and obscured Compton thin (CTN, $10^{22} < N_{\rm H} < 10^{24}\, \rm cm^{-2}$) AGN at different redshift intervals. For clarity, we only plot the SARDs out to redshift $z=2$. The shaded regions in this figure correspond to the 68\% confidence interval estimated from the SARD posterior distributions, while the solid lines represent the median value of the posterior at fixed SAR and redshift intervals. The individual SARD curves in Figure,\ref{fig:sBHAR_z_evol} exhibit a complex shape, broadly characterized by three segments. 
First, a steep-slope regime emerges at high SAR, indicating a rapid decline in the SARD with increasing $\lambda$ beyond a turnover point. 
Second, a plateau where the SARD flattens or mildly drops with increasing $\lambda$, extending to SAR of about 1-1.5\,dex below the turnover point. 
Finally, at very low SAR ($\lambda\approx 10^{-3}-10^{-2}$), the median of the SARD curves decreases with decreasing $\lambda$. This low $\lambda$ regime roughly corresponds to X-ray luminosities $\la 10^{42}\, \rm erg \,s^{-1}$ for the typical stellar masses of the X-ray selected sample, $M_{\star}\approx 10^{10} - 10^{11} \,M_\odot$ (see Figure\,\ref{fig_app:Mstar_lambda_z_distrib}). At these X-ray luminosities, the contribution of normal galaxies to the likelihood in Equation\,\ref{eq:Likelihood_w_gal} becomes significant. Furthermore, in the low $\lambda$ regime, the uncertainties of the SARD measurements increase because of the limited sensitivity of the X-ray selected sample toward low X-ray luminosities.

A striking feature of Figure\,\ref{fig:sBHAR_z_evol} is the redshift evolution of the SARDs, which appear to shift toward higher SAR with increasing redshift. This trend applies to both obscured and unobscured AGN. Therefore the probability of a galaxy hosting a high SAR event increases with redshift, independent of the level of line-of-sight obscuration to the central engine. Additionally, Figure\,\ref{fig:sBHAR_z_evol} reveals a clear difference in the relative normalisation of the unobscured and obscured SARDs at fixed redshift, with the latter being systematically higher. This offset between the two populations reflects the higher fraction of obscured AGN in the Universe, a proportion known to increase with redshift \citep[e.g.][]{La_Franca_2005, Hasinger_2008, Buchner_2015}. Our baseline methodology normalises the SARD via Equation\,\ref{eq:normalisation}, preserving differences in the relative fraction of obscured/unobscured AGN in the results.

Despite the normalisation difference, the similarities between the shape of unobscured and obscured SARDs at fixed redshift in Figure\,\ref{fig:sBHAR_z_evol} are striking. This is further demonstrated in Figure\,\ref{fig:sBHAR_normedNH}. The SARDs in this figure are independently normalised to unity for each obscuration interval to allow comparison of their shapes. Mathematically, this can be expressed by modifying Equation\,\ref{eq:normalisation} as

\begin{equation}\label{eq:normalisation-1}
\int  P(\lambda, N_{\rm H}, z, M_\star) \, d\log\lambda =1.
\end{equation}

\noindent Therefore, the SARDs of Figure\,\ref{fig:sBHAR_normedNH} represent the probability of an AGN in a given obscuration and redshift interval to accrete at a rate\,$\lambda$. Both unobscured and obscured AGN in Figure\ \ref{fig:sBHAR_normedNH} have SARDs with overall shapes that, to the first order, are quite similar. 
However, a closer examination suggests a small ($\approx 1\sigma$ significance) but systematic offset between the SARDs of two populations at fixed redshift, with unobscured extending to higher specific accretion rates. Similar claims for such a trend exist in the literature \citep[e.g.][]{Mountrichas2024}. If this effect is true, it can be interpreted as evidence that fast-accreting black holes are more likely to be unobscured. In Section\,\ref{discussion_orientation}, we discuss further the observed offset and explore systematic uncertainties that could produce it. 

\begin{figure}
    \centering
    \includegraphics[width=0.45\textwidth]{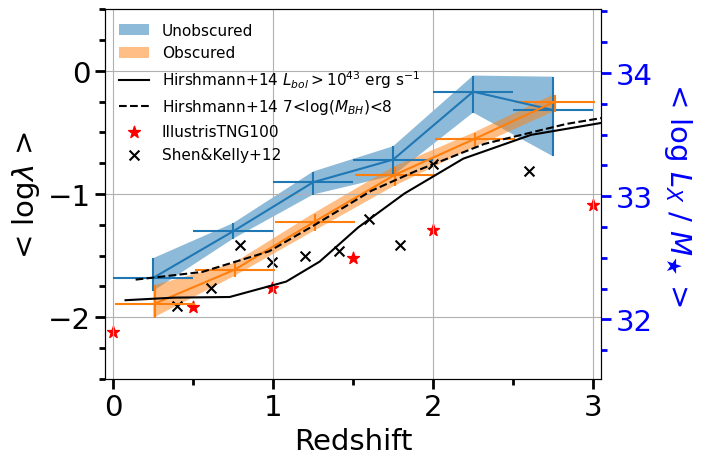}
    \caption{Mean SAR as a function of the redshift for the unobscured ({\it blue}, $\log N_{\rm H}/{\rm cm}^{-2}<22$) and obscured ({\it orange}, $22<\log N_{\rm H}/{\rm cm}^{-2}<24$) populations based on the normalised SARD (Figure\,\ref{fig:sBHAR_normedNH}). The solid lines show the median for each sub-population while the shaded regions correspond to the 1$\sigma$ confidence interval.
    The black solid line is the mean accretion rate from the cosmological hydrodynamic simulations of \protect\cite{Hirschmann_2014} for AGN with bolometric luminosity  $L_{\rm bol}>10^{43}\,{\rm erg\,s}^{-1}$. The black dashed line corresponds to the subpopulation with $7<\log ( M_{\rm BH}/M_{\odot} )<8$ from the same study. The red stars are the Illustris TNG100 cosmological simulation \protect\citep{Habouzit_2019}. The black crosses are the mean Eddington rates of broad line quasars estimated by \protect\cite{Shen_2012}.
    }
    \label{fig:mean_lambda}
\end{figure}

Another way to demonstrate this effect is to explore the redshift evolution of the mean SAR,  <${\rm log}\lambda$>, of the two AGN sub-populations, as shown in Figure\,\ref{fig:mean_lambda}. The averages are estimated from the SARDs of Figure\,\ref{fig:sBHAR_normedNH} for the interval $-3.5\,<\,{\rm log}\lambda\,<\,2$. Unobscured AGN systematically exhibit higher <${\rm log}\lambda$> with respect to obscured ones, up to $z\approx 2.5$. 
Additionally, figure\,\ref{fig:mean_lambda} includes the inferred mean accretion rate evolution of broad line QSOs of \cite{Shen_2012}. These independent estimates for a subpopulation of active SMBHs also show an increasing trend with redshift and are of the same order of magnitude as our results. Also shown in Figure\,\ref{fig:mean_lambda} are the predictions of the cosmological hydrodynamic simulations presented by \cite{Hirschmann_2014} and \cite{Habouzit_2019}. In these models, the mean accretion rate of SMBHs also increases with redshift as a result of the intensity of the adopted AGN (and star-formation) feedback prescriptions that act to remove gas in the nuclear regions of galaxies toward later cosmic epochs. The shallower increasing trend in the case of the Illustris TNG100 simulation \citep{Habouzit_2019} is likely related to the implementation details of the AGN feedback. 

\begin{figure}
    \centering
    \includegraphics[width=0.47\textwidth]{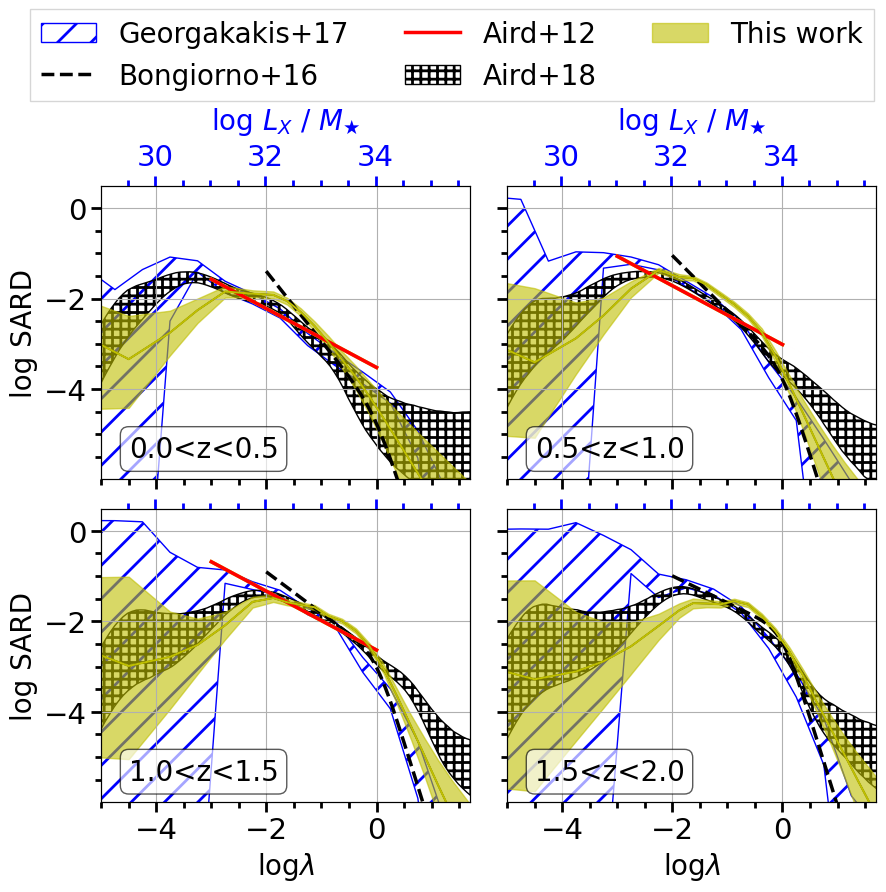}
    \caption{
    The specific accretion rate distribution of CTN AGN ($N_{\rm H}<\rm 10^{24}\,cm^{-2}$) derived in this paper (yellow solid line and shaded region) compared to previous SARD estimates in the literature. Different panels correspond to different redshift intervals. The yellow solid lines represent the median of the SARD posteriors, while the yellow shaded region marks the extent of the 68\% confidence interval around the median. Also overplotted are the results of \protect\cite{Aird_2012} (red solid lines), \protect\cite{Bongiorno_2016} (black dashed lines), \protect\cite{Georgakakis_2017b} (blue hatched areas) and \protect\cite{Aird_2018} (black hatched areas). }
    \label{fig:sBHAR_distrib}
\end{figure}

Figure\,\ref{fig:sBHAR_distrib} provides a further comparison of our SARD measurements with the previous estimates of \cite{Aird_2012}, \cite{Bongiorno_2016}, \cite{Georgakakis_2017b} and \cite{Aird_2018}. We caution that from the studies above, only \cite{Bongiorno_2016} include corrections to account in a statistical manner for the impact of obscuration on the AGN incidence measurements. For this comparison, we add the unobscured and obscured SARDs of Figure\,\ref{fig:sBHAR_z_evol} to obtain the total CTN AGN SARD.  While there is broad agreement in both shape and normalisation between the different SARD estimates presented in Figure\,\ref{fig:sBHAR_distrib}, indicating a consistent emerging picture on the incidence of AGN in galaxies, second-order discrepancies are evident and are attributed to differences in methodology. Our non-parametric SARDs have a systematically higher normalisation at intermediate specific accretion rates $-2\la\log \lambda \la 0$, compared to previous works. This is attributed to the fact that we explicitly estimate obscuration corrections for individual sources and also account for these effects in the X-ray selection function. Also, at low SAR, $\log \lambda\la-2$, our SARD estimates lie below previous estimates. This difference is partly related to the obscuration corrections that act to move obscured AGN to higher intrinsic luminosities and hence, at fixed stellar mass, to higher SAR. Notably, the SARD of \cite{Aird_2018} drops slower toward high $\lambda$ values compared to ours and other estimates in Figure\,\ref{fig:sBHAR_distrib}, reflecting their Bayesian Gamma distribution mixture model methodology.

We also compare our low redshift results with the recent work of \cite{Ananna_2022a, Ananna_2022b} (hereafter \citetalias{Ananna_2022a, Ananna_2022b}) based on the second data release of the BAT AGN spectroscopic survey (BASS DR2). BASS is a low-redshift ($z<0.3$) serendipitous hard X-ray (14-195\,keV) AGN survey, carried out by the \textit{Swift}-BAT telescope \citep{Gehrels_2004, Barthelmy_2005}. This AGN sample is least biased by obscuration because high-energy photons can efficiently penetrate high columns of obscuring material. \citetalias{Ananna_2022a} determine simultaneously the black hole mass function and Eddington-ratio distribution function (ERDF) from the XLF of type\,I and type\,II AGN, defined by the presence or absence of broad Balmer emission lines, respectively. \citetalias{Ananna_2022b} further extend this study to X-ray unobscured ($N_{\rm H}<10^{22}\,{\rm cm}^{-2}$) and obscured ($10^{22}\leq N_{\rm H}<10^{25}\, \rm cm^{-2}$) AGN. It is crucial to note that the ERDF, which represents the space density of AGN in logarithmic Eddington ratio bins, is a distinctly different quantity from the SARD, which expresses the probability of a galaxy hosting an accretion event with a specific accretion rate $\lambda$. Therefore, a direct comparison between the two statistical measures is not straightforward. 
We, therefore, do not display the results of \citetalias{Ananna_2022b} in Figure\,\ref{fig:sBHAR_distrib}, but provide a qualitative comparison with our findings.
In their work, \citetalias{Ananna_2022b} parametrise the ERDF by a broken power-law and infer breaks for unobscured and obscured AGN at Eddington ratios of $-1.06$ and $-1.75$, respectively. These break values are over-plotted as vertical lines in the $z<0.5$ panel of Figure\,\ref{fig:sBHAR_normedNH}. Similarly to our results, \citetalias{Ananna_2022b} find that at high-$\lambda$ unobscured AGN are more likely than their obscured counterparts. Moreover, the ERDF break values above are consistent, at least to the first approximation, with the SAR where the SARDs in Figure\,\ref{fig:sBHAR_normedNH} show a significant change in their slopes. We caution, nevertheless, that the ERDFs of \citetalias{Ananna_2022b} use the Eddington ratio of individual AGN in their sample through direct measurements of their bolometric luminosities and black hole masses \citep{Koss_2022a}. Instead, we use stellar masses as a proxy of the Eddington ratio under the assumption of an underlying  $M_\star-M_{\rm BH}$ correlation. The impact of these assumptions is further discussed in Section\,\ref{subsec:obscured_fraction}.

\subsection{Obscured fraction}\label{subsec:obscured_fraction}

Next, we explore the CTN obscured AGN fraction as a function of the SAR. This is motivated by recent findings (\citetalias{Ricci_2017}, \citetalias{Ricci_2022}) suggesting that this parameter space provides a handle on AGN feedback mechanisms \citep{Alonso_Herrero_2021, Venanzi_2020}.

\begin{figure*}
    \centering
    \includegraphics[width=\textwidth]{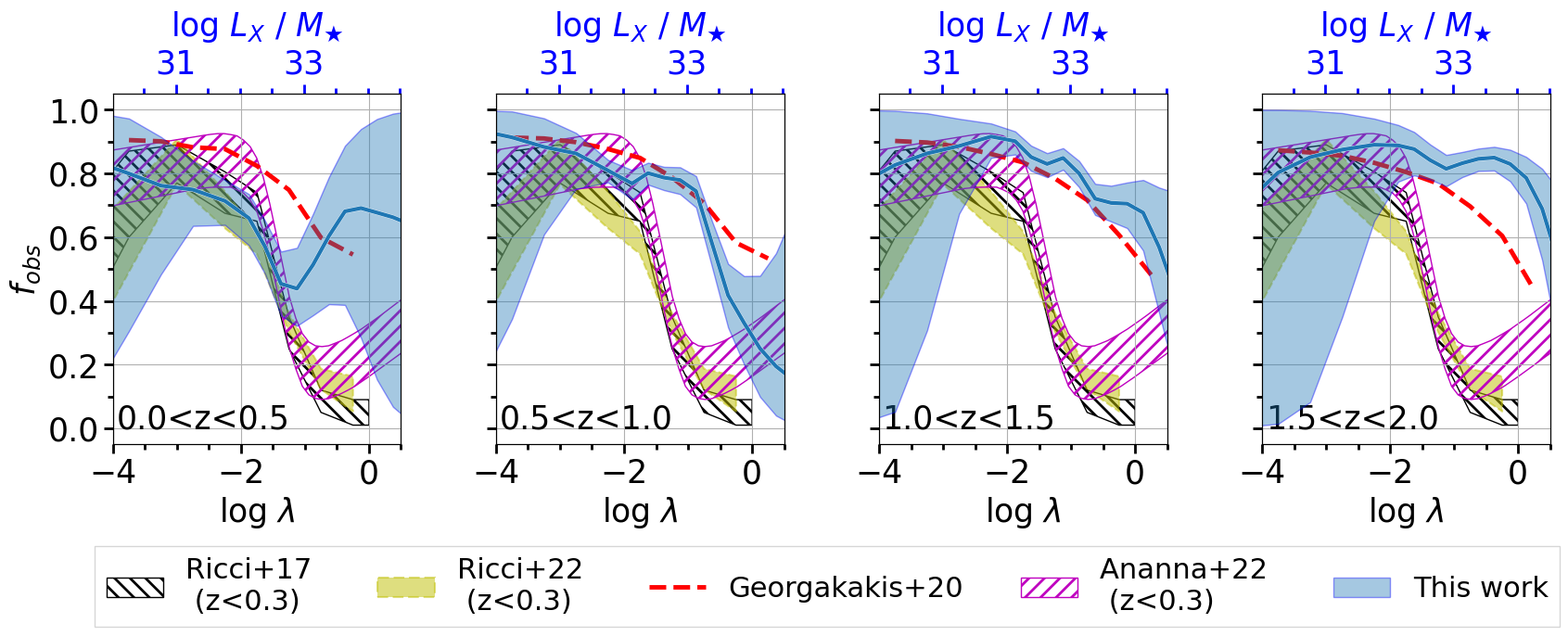}
    \caption{The obscured AGN fraction as a function of SAR $\lambda$. Each panel corresponds to a different redshift interval. The blue solid line represents the median obscuration fraction, while the shaded region corresponds to the 68\% confidence interval around the median. The black-hatched areas are the results of \protect\citetalias{Ricci_2017} in the local Universe (<$z$>= 0.037). The results of \protect\citetalias{Ricci_2022}, based on the BASS DR2 sample, are shown with the yellow shaded area. The purple hatched regions correspond to the observational constraints of \protect\citetalias{Ananna_2022b}. The red dashed curves are the predictions by the forward model with luminosity-dependent obscuration presented in \protect\cite{Georgakakis_2020}.}
    \label{fig:f_obs_z_evolution}
\end{figure*}

Figure\,\ref{fig:f_obs_z_evolution} shows the dependence of the obscured CTN AGN fraction, $f_{\rm obs}$, with specific accretion rate for different redshift intervals up to $z=2$. We obtain $f_{\rm obs}$ by dividing the obscured AGN SARDs presented in Figure\,\ref{fig:sBHAR_z_evol} by the sum of both unobscured and obscured SARDs within each redshift bin. The general trend is that $f_{\rm obs}$ decreases with increasing SAR and this decline becomes steeper toward high $\lambda$ values. The turnover point, $\lambda_{\rm break}$, where the slope changes substantially, appears to evolve with redshift, moving progressively to higher SAR.

At low redshift $z<0.5$, our constraints are in reasonable agreement with the results of (\citetalias{Ricci_2017, Ricci_2022,Ananna_2022b}) up to $\log \lambda \approx -1$. These studies are based on the BASS sample of local AGN \citep{Koss_2017, Koss_2022b}. We find evidence for a significant change of slope at ${\rm log}\,\lambda \approx-2$, consistent with the findings of the studies above. However, the uncertainties in our measurements are important at both very low and high $\lambda$ values due to the limited volume of our sample at low redshift ($z<0.5$).

According to \citetalias{Ricci_2017} and \citetalias{Ricci_2022}, the $\lambda$ dependence on $f_{\rm obs}$ and, more specifically, the steep decline toward high SAR in Figure\,\ref{fig:f_obs_z_evolution} are suggestive of AGN radiation pressure acting on a dusty medium. While the standard Eddington limit, $\lambda=1$, is typically calculated for a fully ionised gas, the AGN obscuring material is known to be partially ionised dusty gas with a cross-section larger than the Thomson one. A larger cross-section induces a lower effective Eddington limit dependent on the column density \citep{Fabian_2008, Fabian_2009}. For a dusty gas at $N_{\rm H}=10^{22}$\,cm$^{-2}$, i.e. the column density limit adopted in this work to separate unobscured from obscured AGN, the effective Eddington limit is $\log \lambda_{\rm eff} \approx -1.7$ \citep{Fabian_2008}. Thus, above this $\lambda$ limit, an AGN effectively accretes at super-Eddington rates and the radiation pressure is expected to blow away the surrounding CTN clouds. As a consequence, the covering factor is reduced, impacting the probability that the AGN is observed as obscured, i.e., the obscured fraction \citep{Ramos_2017}. At $z<0.5$, the results on the obscured AGN fraction plotted in Figure\,\ref{fig:f_obs_z_evolution}, i.e. the sharp decrease of $f_{\rm obs}$ at  $\log \lambda_{\rm eff} \approx -2$, are broadly consistent with this expectation.

At higher redshifts, the turnover point moves to increasingly higher $\lambda$ values, deviating from local Universe results.
This shift could indicate a change in the $\lambda_{\rm eff}$ limit to drive away the obscuring material. For instance, \cite{Fabian_2009} argue that a lower dust grain abundance would increase the low-$N_{\rm H}$ effective Eddington limit. Additionally, the presence of stars in the environment can increase the inward gravitational pull, thereby, shifting the $\lambda_{\rm eff}$ limit towards higher accretion rates. The amplitude of these effects, however, is not expected to be sufficient to explain the shift of the $f_{\rm obs}$ turnover point to high SAR.

The adopted scaling relations to convert the SAR, $L_{\rm X}/M_\star$, to Eddington ratio might also influence the observed trends of the obscured fraction with redshift in Figure\,\ref{fig:f_obs_z_evolution}. For example, a systematic shift towards heavier black hole masses at fixed stellar mass at higher redshifts compared to the adopted local $M_\star-M_{\rm BH}$ relation would lead to a shift of our $f_{\rm obs}$ curves in Figure\,\ref{fig:f_obs_z_evolution} toward lower Eddington ratios. However, studies of the $M_\star-M_{\rm BH}$ relation suggest only mild evolution with redshift \citep{Aversa_2015, Suh_2020, Lopez_2023}, insufficient to explain the 1--2\,dex shift of $\lambda_{\rm break}$ at $z=2$ in Figure\,\ref{fig:f_obs_z_evolution}. 
Also, in our analysis, the bolometric correction used to convert the 2-10\,keV luminosity to bolometric is fixed to $K_{\rm bol}=25$. Instead, it is suggested that $K_{\rm bol}$ increases with increasing luminosity \citep[e.g.][]{Marconi_2003, Duras_2020}. We assess the impact of a variable bolometric correction factor by repeating our Bayesian inference analysis using the \cite{Duras_2020} luminosity-dependent bolometric corrections. We use the new SARD estimates to measure $f_{\rm obs}$ for different redshift intervals, and we find no significant difference with the $f_{\rm obs}$ shapes shown in Figure\,\ref{fig:f_obs_z_evolution}. Furthermore, \cite{Duras_2020} find no dependence of the X-ray bolometric correction on redshift. We therefore conclude that the scaling relations adopted to convert the SAR we measure directly from the observations to Eddington ratio cannot account for the redshift evolution pattern of the $f_{\rm obs}$ in Figure\,\ref{fig:f_obs_z_evolution}.

Studies of the AGN space density evolution and demographics have long been claiming that the obscured AGN fraction depends on AGN luminosity in the sense that obscured AGN become scarcer as the central engine becomes brighter \citep[e.g.][]{Lawrence_1991, La_Franca_2005, Hasinger_2008, Merloni_2014, Ueda_2014, Buchner_2015, Aird_2015, Vijarnwannaluk_2022, Peca_2023}. This trend can be interpreted in the context of the receding torus model, in which the opening angle (i.e. width) of the torus decreases with increasing luminosity \citep{Lawrence_1991}. It is also broadly consistent with the hydrodynamic simulations of the radiation-driven fountain model \citep{Wada_2012}, in which the radiation of the accretion disk drives the gas flow and generates a torus-like structure. It is interesting to assess the extent to which the luminosity dependent-obscuration is consistent with the observations in Figure\,\ref{fig:f_obs_z_evolution} suggesting that $f_{\rm obs}$ is dependent of $\lambda$. For this exercise, we use the empirical model of \cite{Georgakakis_2020}. They infer the SARD of AGN by assuming that convolving it with the galaxy SMF at a given redshift reproduces the X-ray luminosity function of AGN. \cite{Georgakakis_2020} use the \cite{Ilbert_2013} SMF and the \cite{Aird_2015} X-ray luminosity function (including obscured AGN) to solve the above inverse convolution problem and estimate a parametric model for the SARD of AGN as a function of redshift. This SARD parametrisation is then used to assign specific accretion rates to mock galaxies drawn from the SMF and hence, produce simulated AGN samples. These AGN are further assigned column densities,  $N_{\rm H}$, in a probabilistic way based on the obscuration model of \cite{Aird_2015}. Inherent in these mock AGN catalogues is the X-ray luminosity-dependent obscuration inferred by \cite{Aird_2015}. We use the \cite{Georgakakis_2020} empirical model to make predictions on how $f_{\rm obs}$ varies as a function of specific accretion rate and to overplot the results in Figure\,\ref{fig:f_obs_z_evolution}. This rather simple empirical model performs reasonably well and, to the first approximation, is consistent with the observations in Figure\,\ref{fig:f_obs_z_evolution} especially at $0.5<z<1.5$. In detail, however, there are discrepancies suggesting that AGN luminosity may not be the only parameter that affects the covering factor of the obscuring material in AGN.

\subsection{Evolution of the blow-out region}\label{subsec:blowout_region}

\begin{figure}
    \centering
    \includegraphics[width=0.45\textwidth]{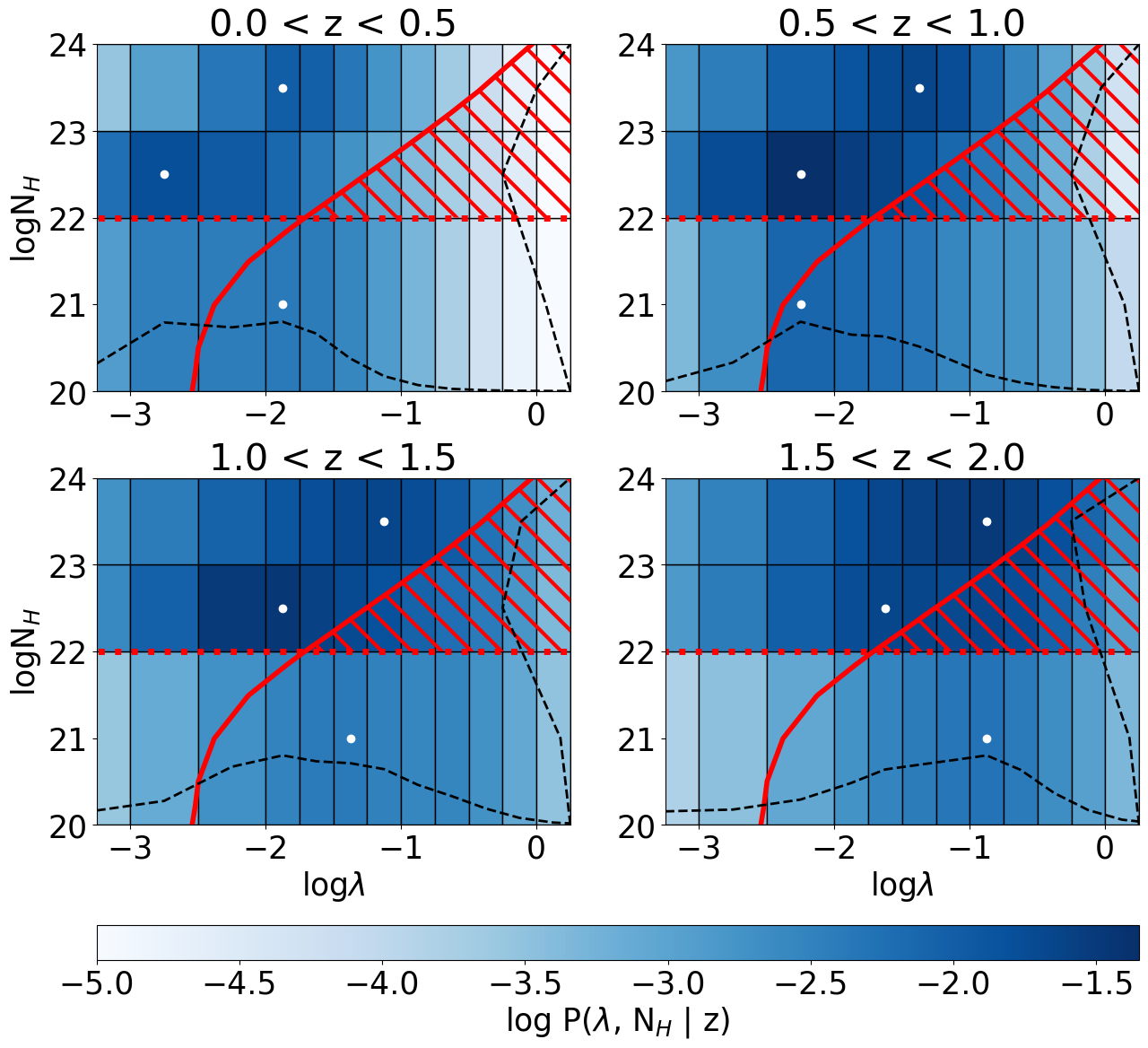}
    \caption{Logarithmic probability density $\log P(\lambda, N_{\rm H} | z)$ of a galaxy to host an AGN as a function of SAR (log$\lambda$) and obscuration (log$N_{\rm H}$). Each panel represents a different redshift interval from 0 to 2. The diagonal solid red line represents the effective Eddington limit as a function of the column density \protect\citep{Fabian_2008} while the horizontal dashed line shows the expected contribution of the host galaxy to the obscuration level. The red-hatched area in between the two curves is the blowout region, an unstable and short-lived region where the radiation pressure should push away the gas efficiently.  The white dots indicate, for each redshift and log$N_{\rm H}$ interval, the $\lambda$-bin (log$\lambda>-3.5$) with the highest probability. At the bottom of each panel, the horizontal dashed black curve corresponds to the probability density summed over log$N_{\rm H}$ as a function of log$\lambda$. Respectively, the  vertical dashed black curve is the probability density summed over log$\lambda$ as a function of log$N_{\rm H}$.}
    \label{fig:blowout_region_evol}
\end{figure}

Figure\,\ref{fig:blowout_region_evol} provides an alternative visualization of the obscured fraction evolution, depicting the SARD as a function of both $N_{\rm H}$ and $\lambda$. The shading of each cell represents the probability density for an AGN of being in that cell at a given redshift and with a stellar mass within $9.5 < \log(M_\star /M_\odot)< 11.5$. It reflects the intrinsic and unbiased distribution of AGN in this plane. The probability values shown in this figure are computed using the same methodology described in Section\,\ref{subsec:sBHAR_calculation}, except that the boundaries of the ${\rm log}(N_{\rm H}/{\rm cm}^{-2})$ grid are now set at [20, 22, 23, 24].
In each panel of Figure\,\ref{fig:blowout_region_evol}, the red solid line corresponds to the effective Eddington limit for a dusty gas as a function of the column density \citep{Fabian_2008}. 
The region located above the obscuration limit $N_{\rm H}>10^{22}$cm$^{-2}$ and on the right of the effective Eddington limit curve is referred to as the blow-out or {\it forbidden} region \citep{Venanzi_2020}. As mentioned previously, due to the dustiness of the surrounding gas, sources in this region are effectively accreting at super-Eddington rates despite having $\lambda < 1$. The radiation pressure of these AGN can potentially launch winds that clear up the obscuring material in a short time scale relative to the overall AGN phase. Therefore, due to being short-lived and unstable, the fraction of AGN detected in that region is expected to be small hence, its name \citep{Fabian_2009, Ricci_2017, Ricci_2022}.

In the $z<0.5$ panel of Figure\,\ref{fig:blowout_region_evol}, the blow-out region is sparsely populated, consistent with the predictions of the theoretical picture presented above. However, with increasing redshift, the {\it forbidden} region appears progressively more populated.
The white dots in each panel represent the peaks of the AGN probability distribution for each $N_{\rm H}$ interval and they shift towards higher SAR with increasing redshift. This global increase of the accretion activity of AGN with redshift is better visualised by the probability summed over $N_{\rm H}$ displayed with a dashed line at the bottom of each panel in Figure\,\ref{fig:blowout_region_evol}. 
Conversely, the probability density summed over $\lambda$, represented by the dashed line on the right side of each panel, shows an increase of obscured AGN with redshift.

\begin{figure}
    \centering
    \includegraphics[width=0.45\textwidth]{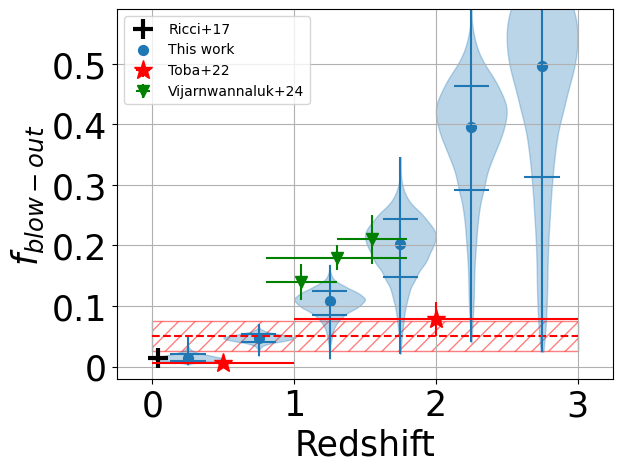}
    \caption{Fraction of sources within the {\it forbidden} region as a function of the redshift. The violin plots represent the posterior distribution of our $f_{\rm blow-out}$ value for each redshift interval, and the blue dots and error bars correspond to the median values and $1\sigma$-uncertainties, respectively.
    The {\it black} cross corresponds to the value presented in \protect\cite{Ricci_2017}. 
    The {\it red} crosses are the results of \protect\cite{Toba_2022} within the $z=0-1$ and $z=1-3$ redshift intervals, using soft X-ray data from {\it eROSITA}. Through a Monte Carlo randomization, \protect\cite{Toba_2022} find a constant value of $5\pm 2.5\%$ represented by the {\it red hatched} area. 
    The green triangles are the $f_{\rm blow-out}$ values from \protect\cite{Vijarnwannaluk_2024}.
    }
    \label{fig:f_blowout_z_evolution}
\end{figure}

From Figure\,\ref{fig:blowout_region_evol}, it is possible to estimate the fraction of AGN within the blow-out region as

\begin{equation}
f_{\rm blow-out} = \frac{\sum\limits_{(\lambda, N_{\rm H}) \;\in \;\mathcal{R}_f}^{} P(\lambda, \, N_{\rm H} \,| \,z)} { \sum\limits_{\log\lambda>-3.5} P(\lambda, \, N_{\rm H} \,| \,z) }, 
\end{equation}\label{eq:f_blowout}

\noindent where $P(\lambda, \, N_{\rm H} \,| \,z)$ is the probability in the two dimensional space ($\lambda, \, N_{\rm H}$) of Figure\,\ref{fig:blowout_region_evol} and $\mathcal{R}_f$ is {\it forbidden} region of the parameter space. For the summation of the numerator, we attribute to the cells that partially overlap with the blow-out region a weight corresponding to the fraction of their area within that region. 
The summation of the denominator is over all cells with $\log\lambda>-3.5$ and represents the AGN duty-cycle, i.e. the probability that a galaxy hosts an AGN with $\log\lambda$ higher than the threshold $-3.5$. The dependence of the duty-cycle on the chosen $\lambda$-threshold for different redshifts is presented in Figure\,\ref{fig_app:duty_cycle} in the Appendix.
The choice of the $\lambda$-threshold affects the width of the uncertainties but does not impact our results and conclusions.

The evolution of $f_{\rm blow-out}$ with redshift is shown in Figure\,\ref{fig:f_blowout_z_evolution} in the form of violins representing the posterior distribution for each redshift interval. The median $f_{\rm blow-out}$ values increase with the increasing redshift with over 10\% of the AGN at $z>1$ being found in the {\it forbidden} region. This significant evolution of $f_{\rm blow-out}$ is at odds with the very definition of the {\it forbidden} region. Nevertheless, we caution that the uncertainties of $f_{\rm blow-out}$ also progressively rise with redshift, becoming significant at $z>1.5$. These large error bars reflect the increasing uncertainty in the determination of the AGN duty cycle, $\sum_{\log\lambda>-3.5} P(\lambda, \, N_{\rm H} \,| \,z)$, i.e. the denominator of Equation\,\ref{eq:f_blowout}. For instance, within the redshift range $1.5<z<2$, the median duty cycle and $1\sigma$ errors are $12^{+4}_{-2}\%$, but the tails of the posterior distribution are pronounced with the 3\,$\sigma$ upper confidence limit being 70\%. Consequently, the posterior distribution of $f_{\rm blow-out}$ also has an extended posterior tail to lower values.

In Figure\,\ref{fig:f_blowout_z_evolution}, we also compare our results with the local Universe measurement of \citetalias{Ricci_2017} based on the BASS sample. They estimate $f_{\rm blow-out}=1.4\%$ in good agreement with our measurement at $z<0.5$, $f_{\rm blow-out}=1.4^{+0.6}_{-0.4}\%$. Additionally, we include the findings of \cite{Toba_2022} in this figure, who investigated the multiwavelength properties of 692 X-ray selected AGN with WISE W4 counterparts detected in the {\it eROSITA} Final Equatorial-Depth Survey \citep[eFEDS,][]{Brunner_2022}. Under the assumption of a constant AGN blow-out fraction in the redshift interval $z=0-3$, they infer $f_{\rm blow-out}=5 \pm 2.5\%$, illustrated by the red shaded region in Figure\,\ref{fig:f_blowout_z_evolution}. Our results, however, suggest an increasing trend for $f_{\rm blow-out}$ toward high redshift, contradicting the constant fraction assumption of \cite{Toba_2022}. Nevertheless, when these authors split their sample into low ($z=0-1$) and high ($z=1-3$) redshift bins, they find a mild increase. It is worth underlying that {\it eROSITA} is a soft X-ray (0.3-2.3\,keV) telescope \citep{Liu_2022} and, therefore, biased against obscured systems.
Additionally, Figure\,\ref{fig:f_blowout_z_evolution} presents the recent results of \cite{Vijarnwannaluk_2024} which, by exploiting the HSC-DEEP {\it XMM}-LSS survey data, find $f_{\rm blow-out} = 0.18 \pm 0.02$ in the redshift interval $0.8 < z < 1.8$. When splitting the redshift interval in half, an $f_{\rm blow-out}$ increasing trend similar to our results is noticeable with $f_{\rm blow-out} = 0.14 \pm 0.03$ at $0.8 < z < 1.3$ and $f_{\rm blow-out} = 0.21 \pm 0.04$ at $1.3 < z < 1.8$.

The adopted scaling relations to convert the specific accretion rate to Eddington ratio (see Equation\,\ref{eq:sar2edd}) clearly affects the inferred fraction of AGN within the {\it forbidden} region of Figure\,\ref{fig:blowout_region_evol}. For example, if the $M_{BH}-M_{\star}$ relation evolves with redshift in a way that leads to more massive BHs at fixed stellar mass, then the net effect would be an overall shift of the probabilities of individual cells in Figure\,\ref{fig:blowout_region_evol} toward lower Eddington ratios. This would reduce the estimated fraction of AGN in the blowout region in Figure\,\ref{fig:f_blowout_z_evolution}. However, as discussed in Section\,\ref{subsec:obscured_fraction}, studies of the $M_{BH}-M_{\star}$ relation suggest little evidence for redshift evolution. Introducing scatter in $M_{BH}-M_{\star}$ relation would tend to blur the estimated probabilities of individual cells in Figure\,\ref{fig:blowout_region_evol} by distributing them to nearby $\lambda$ bins but is not expected to drastically modify the observed trends. The choice of the X-ray bolometric correction also has little impact on the results shown in Figures \ref{fig:blowout_region_evol} and \ref{fig:f_blowout_z_evolution}. 
The increase of $f_{\rm blow-out}$ could potentially indicate a decrease in the efficiency of the dusty radiation pressure mechanism to launch outflows. Varying the dust grain abundance, for instance, reduces the area of the {\it forbidden} region \citep{Fabian_2009}, resulting in lower $f_{\rm blow-out}$ values. However, changing the abundance from 1 to 0.3 or 0.1 does not change the behaviour of the $f_{\rm blow-out}$ posterior distribution that remains increasing with redshift. The only difference is that the median values are slightly lower compared to those plotted in Figure\,\ref{fig:f_blowout_z_evolution}, with a maximum difference of $\sim 4\%$ at high redshift.

\section{Discussion}\label{Sec:Discussion}

This paper presents the first quantitative analysis of how AGN with varying levels of line-of-sight obscuration inhabit galaxies across cosmic time, from the local Universe to high redshifts, $z\sim3$. We combine three extragalactic X-ray survey fields (COSMOS, AEGIS-XD, CDFS) with distinct X-ray depths and sizes on the sky to compile a sample of 3882 X-ray selected AGN with 2-10\,keV luminosities in the range $10^{40}-10^{46} \rm \,erg\,s^{-1}$ and redshifts spanning $z=0.01-5$ (see Figure\,\ref{fig:Lx_vs_z}). 
We employ state-of-the-art Bayesian X-ray analysis techniques to constrain the intrinsic 2-10\,keV X-ray luminosity, $L_{\rm X}$, and hydrogen line-of-sight column density, $N_{\rm H}$, of individual sources in the sample. 
In parallel, template fitting of the observed UV-to-mid-IR SEDs yields constraints of the stellar masses $M_\star$ of the X-ray AGN host galaxies. Combining these observables enables us to derive the specific accretion rate $\lambda \propto L_{\rm X}/M_\star$ and its associated uncertainty for each source. We develop a Bayesian framework that accounts for selection effects and observation uncertainties to derive the SARD, i.e. the probability of a galaxy at a given redshift hosting an AGN with a given $\lambda$ and $N_{\rm H}$.
In the following sections, we discuss the redshift evolution of the SARD (Section\,\ref{discussion_redshift}) and how our findings provide information on the origin of the obscurer (Section\,\ref{discussion_orientation} and\,\ref{discussion_ISM}).

\subsection{Redshift evolution of SARD: evidence for downsizing?}\label{discussion_redshift}

\begin{figure}
    \centering
    \includegraphics[width=0.45\textwidth]{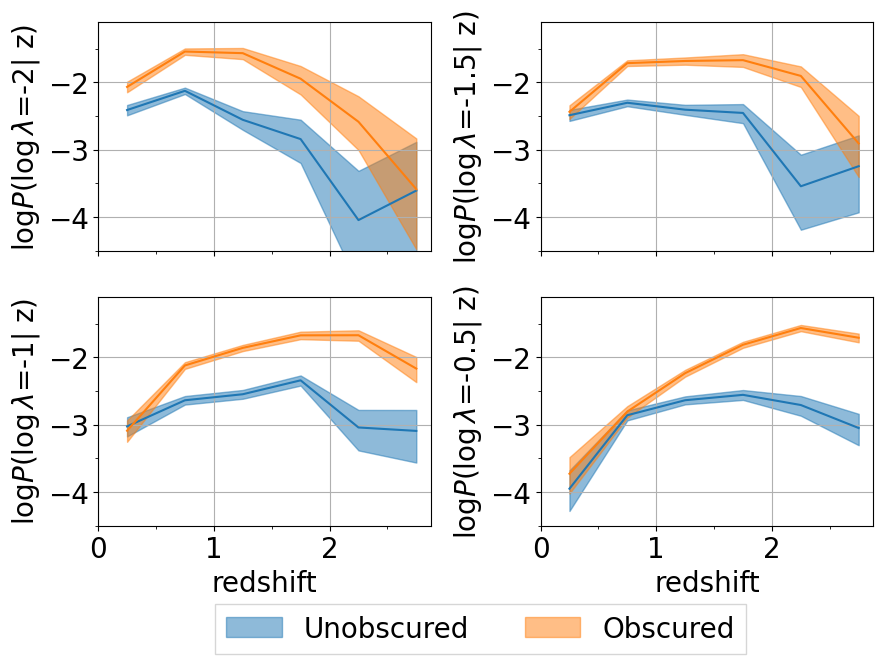}
    \caption{Redshift evolution of the probability of an AGN to accrete at a fixed SAR $\log\lambda$ for different $\lambda$ values. It is a different way to show the redshift evolution of the SARD presented in Figure\,\ref{fig:sBHAR_z_evol}.
    The {\it blue} and {\it orange shaded} regions represent the probability when normalised over $N_{\rm H}$ for unobscured ($\log N_{\rm H}/{\rm cm}^{-2}<22$) and obscured ($22<\log N_{\rm H}/{\rm cm}^{-2}<24$) AGN, respectively.}
    \label{fig:z_evol_at_fixed_lambda}
\end{figure}

Our analysis shows that the specific accretion rate of both unobscured ($20<{\rm log}(N_{\rm H}/{\rm cm}^{-2})<22$) and obscured ($22<{\rm log}(N_{\rm H}/{\rm cm}^{-2})<24$) AGN has a complex shape that can be roughly approximated by a power-law, bending to a steeper slope at high SARs, and showing evidence for a turnover at the low-SAR end (see Figure\,\ref{fig:sBHAR_z_evol}). Interestingly, the high SAR bending point, where the slope of the distribution changes significantly, appears to shift to higher specific accretion rates with increasing redshift. This evolution pattern applies to both the obscured and unobscured AGN population and results in an increase of the mean SAR with increasing redshift, e.g. Figure\,\ref{fig:mean_lambda}. Similar results have been reported in previous studies \citep[e.g.][]{Shen_2012, Georgakakis_2017b} and align with the predictions of cosmological models on the evolution of galaxies and their supermassive black holes \citep[e.g.][]{Sijacki_2015, Weinberger_2018}. 
The physical interpretation of this trend is tied to the impact of AGN feedback on the gas availability in the vicinity of SMBHs \citep[e.g.][]{Hirschmann_2014, Habouzit_2019}. 

Another perspective on the SARD evolution pattern is presented in Figure\,\ref{fig:z_evol_at_fixed_lambda}, which plots the probability of a galaxy hosting an AGN at fixed SAR as a function of redshift. With increasing SAR (different panels in Figure\,\ref{fig:z_evol_at_fixed_lambda}), the peak or plateau of the corresponding curves shift to higher redshift, with the trend being more pronounced in the case of obscured AGN. Figure\,\ref{fig:z_evol_at_fixed_lambda} is reminiscent of the concept of AGN "downsizing" \citep{Barger_2005,
Hasinger_2005, Shankar_2013}, originally defined as the fact that the accretion density of low-luminosity systems peaks at lower redshift compared to luminous AGN.  
Previous studies indeed suggest a connection between AGN downswing and the evolution of the incidence of AGN in galaxies. \cite{Georgakakis_2017b} for example, find evidence that at earlier times, the relative probability of high versus low specific accretion-rate events among galaxies increases, in qualitative agreement with the trends in Figure\,\ref{fig:z_evol_at_fixed_lambda}. They argue that this differential redshift evolution of the AGN duty cycle with respect to SAR produces the AGN downsizing trend. Similarly, \cite{Bongiorno_2016} argue that the differential redshift evolution of the AGN SARD compared to the weak evolution of the host galaxy SMF are the main factors behind the AGN downswing trend, which is reflected by the evolution of the SARD, i.e. shift to higher SAR with increasing redshift, at fixed $M_\star$ range.
Moreover, our analysis adds to the studies above by showing that SARD evolution applies separately to both obscured and unobscured AGN. Therefore, the underlying physical mechanism driving this trend applies nearly uniformly to all AGN, regardless of obscuration level. The well-established increase in the gas content of galaxies to $z \sim 3$ \citep{Santini_2014, Tacconi_2020} likely plays an important role in establishing the patterns shown in Figures\,\ref{fig:sBHAR_z_evol}, \ref{fig:mean_lambda} and\,\ref{fig:z_evol_at_fixed_lambda}.

\subsection{Origin of the obscurer: Support of a modified orientation model}\label{discussion_orientation}

In this subsection, we discuss the results of the comparison of the SARDs of obscured and unobscured AGN, contextualizing them within models explaining the nature of the obscurer. The simplest version of the orientation or unification scenario \citep{Antonucci_1993, Urry_1995} postulates that all AGN are surrounded by obscuring material with a toroidal shape \citep{Ramos_2017, Combes_2019, Garcia_Burillo_2019, Garcia_Burillo_2021, Gamez_Rosas_2022} and a universal opening angle. The obscuration properties of individual AGN are determined by geometrical effects, i.e. if the line of sight to the observer intersects the obscurer. In this picture, since obscured and unobscured AGN differ only in their orientation relative to the observer, their SARDs are expected to have similar shapes. Figure\,\ref{fig:sBHAR_z_evol} is consistent with this scenario, showing that, to the first approximation, the SARDs of AGN samples categorized by hydrogen column density $N_{\rm H}$ are similar once differences in their normalisations are factored out.

In detail, however, there is evidence in Figure\,\ref{fig:sBHAR_normedNH} for subtle differences between the SARDs of obscured and unobscured AGN, in the sense that the SARD high-$\lambda$ turnover point of unobscured systems is located towards higher SAR compared to obscured AGN. 
As illustrated in Figure\,\ref{fig:f_obs_z_evolution}, this offset leads to a fast decline in the obscured AGN fraction, directly related to the opening angle of the dust and gas clouds surrounding the central engine. This finding aligns with studies on the Eddington ratio distribution of type\,I and type\,II AGN in the local Universe (\citetalias{Ricci_2017, Ricci_2022, Ananna_2022b}). 
Nevertheless, we caution against potential systematic and random errors that might affect the observed trend, such as biases in determining the stellar mass of galaxies hosting AGN. In particular, for type\,I AGN, estimating host galaxy properties by fitting templates to the observed SEDs is non-trivial \citep{Vanden_Berk_2001, Richards_2006} and requires the decomposition of the stellar emission from the AGN light. 
For example, an ad-hoc systematic stellar mass bias of +0.3\,dex for unobscured AGN suffices to mask any differences between the SARDs of obscured and unobscured AGN in Figure\,\ref{fig:sBHAR_z_evol}. However, previous studies either do not find systematic differences in the stellar mass distribution of obscured and unobscured AGN \citep{Andonie_2022} or suggest that obscured AGN are actually more massive than unobscured ones \citep{Mountrichas_2021, Mountrichas_2023}.
Additionally, the $M_{\rm BH}-M_\star$ scaling relation used in our study for converting SAR to Eddington ratio, may impact our results due to its scatter and potential dependence on obscuration \citep{F_Ricci_2017}.
Nevertheless, despite these potential caveats, our results indicate a second-order but significant difference in the accretion properties of unobscured and obscured AGN.

The dependence of the obscured AGN fraction on physical parameters, such as accretion luminosity or Eddington ratio, provides additional information on the origin of the obscuration in AGN. We explore this potential in Figure\,\ref{fig:f_obs_z_evolution}, which plots the obscured AGN fraction as a function of $\lambda$. The observed trends in this figure broadly align with a modified version of the unification model, suggesting a reduction in the opening angle of the obscurer as luminosities increase \citep[e.g.][see discussion in Section\,\ref{subsec:obscured_fraction}]{Lawrence_1991}. An alternative scenario for the dependence of the obscured fraction on SAR is the radiation-regulated model (\citetalias{Ricci_2017, Ricci_2022}) that links AGN obscuration to distinct stages of the growth of black holes. In this picture, AGN feedback acting on dusty gas clouds \citep{Fabian_2008, Fabian_2009} regulates the evolution of AGN on the $\lambda - N_{\rm H}$ plane of Figure\,\ref{fig:blowout_region_evol}. 
Initially, an accretion event onto an SMBH is triggered by some process (e.g. disk instabilities, galaxy interactions). The quiescent black hole starts off at the far left end of  $\lambda-N_{\rm H}$ diagram of Figure\,\ref{fig:blowout_region_evol} and moves gradually to higher $\lambda$. At the same time, the line-of-sight obscuration of the system increases as a result of the inflowing material of gas and dust clouds. As a result, the AGN moves toward the top left corner of the $\lambda-N_{\rm H}$ plane, where the obscurer has a larger covering factor, e.g. $\sim 85\%$ at log$\lambda<-2$ in the work of \citetalias{Ricci_2017}. As the Eddington ratio continues to increase, the AGN enters the {\it forbidden} region of the parameter space, where radiation pressure acting on the dust grains renders the system unstable to outflows \citep{Fabian_2008, Fabian_2009} that can act to gradually push away the CTN obscuring material. The final stage of black hole growth consists of a rapidly accreting unobscured AGN with a low covering factor at the bottom right corner of the diagram. 
Ultimately, as the gas and dust are consumed by the black hole or blown away, the accretion rate gradually decreases, and the system returns to its initial quiescent state.

This modified orientation scenario that includes feedback predicts a decrease in the obscured AGN fraction beyond $\log \lambda \approx -2$, corresponding to the effective Eddington limit of dust clouds with $N_{\rm H}\approx 10^{22}\, \rm cm^{-2}$. As explained in Section\,\ref{subsec:blowout_region}, this obscured AGN fraction decline is attributed to the existence of the {\it forbidden} region ({\it red-hatched} area in Figure\,\ref{fig:blowout_region_evol}), in which the incidence of AGN is theoretically very low. Our results in the smallest redshift bin ($z<0.5$) suggest that $f_{\rm blow-out}\sim 1\%$, in agreement with local Universe studies (\citetalias{Ricci_2017}). 
However, as illustrated in Figure\,\ref{fig:f_blowout_z_evolution}, the fraction of AGN in this region increases significantly with redshift, exceeding 30\% of AGN at $z>2$, surpassing previous estimations \citep{Toba_2022}. 
Consequently, as depicted in Figure\,\ref{fig:f_obs_z_evolution}, $\lambda_{\rm break}$ at which the obscured AGN fraction declines shifts with redshift, from $\log\lambda\sim -2$ to higher SAR, no longer consistent with the radiation-regulated model predictions.
The nature of the AGN driving this redshift evolution of the {\it forbidden} region is key to understand the intrinsic relation between obscuration and accretion.

To start with, it is important to acknowledge that the effective Eddington limit, bordering the blow-out region, is not universal. Its characteristics depend on physical parameters that can differ among various AGN \citep{Arakawa_2022}. \cite{Fabian_2009} demonstrate that the dust-to-gas ratio can impact the shape of this limit, while the presence of stars can shift the $\lambda_{\rm eff}$ limit to higher accretion rates. Expanding on this, \cite{Ishibashi_2018} suggest that radiation trapping makes the high-obscuration end of the $\lambda_{\rm eff}$ limit more vertical, expanding the size of the {\it forbidden} region. Furthermore, they propose that factors such as the inclination angle of the line-of-sight or external gas pressure due to gas inflow can shift the $\lambda_{\rm eff}$ limit along the $\lambda$-axis. Additional assumptions, like the isotropy of the emission, can also impact the effective Eddington limit. While these physical dependencies may contribute to the scatter of AGN in the $\lambda-N_{\rm H}$ plane at low redshifts ($z<0.5$), they appear too weak to account for the observed redshift evolution.

By definition, AGN in the blow-out region should exhibit outflows that push away the nuclear-scale obscuring material through radiation pressure \citep{Fabian_2008, Fabian_2009}. 
For instance, \cite{Kakkad_2016} find that 4 out of the 5 AGN selected in the {\it forbidden} region show clear outflow signatures.
Similarly, using the eFEDs catalogue \citep{Brunner_2022}, \cite{Musiimenta_2023} compare the outflow properties of mid-IR colour-selected AGN \citep[sample A;][]{Brusa_2015, Zakamska_2016, Perrotta_2019} with AGN selected within the blow-out region (sample B).
Remarkably, sample B contains 528 sources ($\sim 14\%$ of the parent eFEDS population with optical spectra at $z>0.5$), reinforcing the idea of a non-empty {\it forbidden} region.
Additionally, both samples A and B exhibit high outflow incidence (3 out of 4 in sample B within the redshift of interval and with good enough spectroscopy). However, sample B shows bluer colours than sample A, which could suggest the initial stage of the blowout phase or the presence of inhomogeneous obscuration.
These results are supported by the work of \cite{Yamada_2021}, which suggests that chaotic quasi-spherical inflows in late mergers produce dusty outflows, responsible for the obscuration in the blow-out region.
If the expected high incidence of outflows in {\it forbidden} region AGN is confirmed in larger samples, thereby, the increasing $f_{\rm blow-out}$ with redshift could indicate an increased incidence of outflows in the overall AGN population with redshift. However, previous studies claim that the fraction of [O III] outflows in AGN does not significantly evolve with redshift \citep{Harrison_2016, Coatman_2019}.

\subsection{Origin of the obscurer: Evidence for ISM obscuration}\label{discussion_ISM}

The redshift evolution of $f_{\rm obs}$ and $f_{\rm blow-out}$ can also be explained by a different origin of the obscuration. According to evolutionary models \citep[e.g.,][]{Sanders_1988, Di_matteo_2005, Hopkins_2006, Klindt_2019}, obscuration is linked to a specific stage in the life cycle of an AGN. During this stage, infalling gas and dust that fuel the SMBH create a cocoon of obscuring material around it \citep{Lapi_2018, Sicilia_2022}. This connection ties AGN growth and feedback activity to host galaxy properties, such as the star formation rate \citep{Lapi_2014}.
The main distinction from the earlier mentioned radiation-regulated model (\citetalias{Ricci_2017}) is that the obscurer is not confined to the nuclear scale. It can also originate from the interaction of the photons with the interstellar medium (ISM) of the host galaxy on larger scales.
A direct consequence of this large-scale ISM obscuration is a longer time-scale for the feedback to clear out the ISM obscuring material \citep[e.g. $\rm 10^{6-7}\,yr$ for distances $\rm 10^{2-3}\,pc$ from the central engine;][]{King_2015, Hyunsung_2021} compared to nuclear-scale obscuration \citep[e.g. $\rm \approx 2\times 10^{5}\,yr$ for distances of about 30\,pc;][]{Lansbury_2020, Hyunsung_2021}. Therefore, ISM-obscured AGN can have a higher incidence in the blow-out region compared to torus-obscured AGN.
The powerful dust-reddened quasars are believed to be obscured by the ISM \citep{Alexander_2012, Klindt_2019, Andonie_2022}, and represent a younger, transitioning phase than their bluer counterparts \citep{Yi_2022}. Interestingly, previous studies found a significant number of such red quasars within the {\it forbidden} region \citep{Glikman_2017, Lansbury_2020}. Furthermore, in a recent study, \cite{Glikman_2024} examine a sample of 10 red quasars from the FIRST-2MASS survey and noticed that all quasars hosted in galaxies exhibiting merging signatures were found within the blow-out region. These findings suggest the importance of the host galaxy evolution on the AGN properties.

However, a limitation of this work is the difficulty of disentangling nuclear-scale and large-scale obscuration from X-ray spectroscopy alone. The ISM contribution to the line-of-sight obscuration $N_{\rm H}$ is considered insignificant at low redshift. Nevertheless, several studies point to an increasing ISM contribution with higher redshifts \citep{Buchner_2017a, Gilli_2022, Alonso_Tetilla_2023}, which can even dominate in compact starburst galaxies \citep{Andonie_2023}. For instance, \cite{Buchner_2017a} use gamma-ray bursts to probe the host galaxy obscuration and find a $M_\star-N_{\rm H,\, ISM}$ relation suggesting a higher ISM obscuration contribution in the most massive galaxies. Similarly, \cite{Buchner_2017b} observe an ISM obscuration increase with redshift.
Moreover, \cite{Gilli_2022} estimate that the median ISM surface density evolves with $(1+z)^{3.3}$, suggesting that the ISM obscuration could even exceed the CTK limit at $z \geq 6$. 
Additionally, the recent work of \cite{Andonie_2023} uses a similar methodology as \cite{Circosta_2019}, independently measuring the size of host galaxies with ALMA observations and the total gas mass in them to constrain ISM obscuration. By doing so, \cite{Andonie_2023} find a dependence between obscuration and star formation rate and observationally confirm that ISM obscuration can even be CTK in the most compact starburst galaxies.

\begin{figure}
    \centering
    \includegraphics[width=0.99\linewidth]{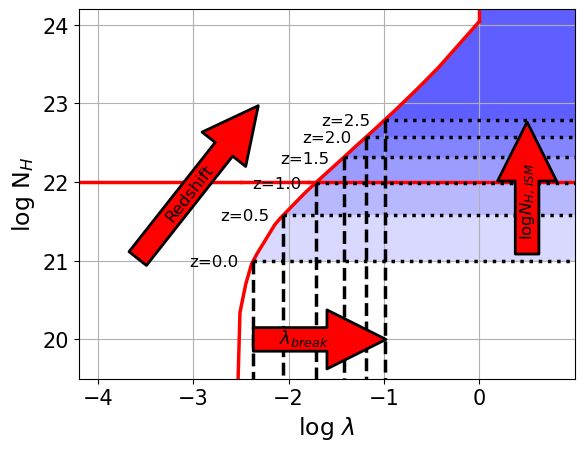}
    \caption{Schematic redshift evolution of the impact of the ISM obscuration on the {\it forbidden} region across the $\lambda-N_{\rm H}$ plane. 
    The diagonal {\it red solid} line corresponds to the effective Eddington limit \protect\citep{Fabian_2008} while the {\it horizontal red solid} line corresponds to the $N_{\rm H}=10^{22}\,{\rm cm}^{-2}$ threshold \protect\citep{Matt_2000} delimiting the {\it forbidden} region in this work. 
    With increasing redshift, the median host galaxy ISM obscuration increases following $N_{\rm H,\,ISM}=10^{21} \times (1+z)^{3.3}\,{\rm cm}^{-2}$ \protect\citep[{\it horizontal black dotted} lines; ][]{Gilli_2022}. 
    As a consequence, $\lambda_{\rm break}$, at which the ISM obscuration and the effective Eddington limit intersect, shifts towards higher SAR ({\it vertical black dashed} lines). $\lambda_{\rm break}$ corresponds to the SAR at which the obscured AGN fraction drops.
    }
    \label{fig:sketch_forbidden_evolution}
\end{figure}

The impact of the increasing ISM obscuration with redshift on the {\it forbidden} region is illustrated in Figure\,\ref{fig:sketch_forbidden_evolution}. In our study, the obscuration lower limit of the blow-out region is fixed to $N_{\rm H}=10^{22}\,{\rm cm}^{-2}$ \citep{Matt_2000}. Above this threshold, we assume that the obscuration is only attributed to the torus. 
However, according to the findings of \cite{Gilli_2022}, the ISM obscuration is increasing with redshift (vertical large {\it red arrow} in Figure\,\ref{fig:sketch_forbidden_evolution}) and the median ISM obscuration, $N_{\rm H,\,ISM}$, is larger than the $10^{22}\,{\rm cm}^{-2}$ threshold at $z>1$.
Therefore, at higher redshifts, an AGN in the {\it forbidden} region would have its nuclear-scale obscurer rapidly blown away as in the local Universe, but, due to longer clearing time scales, the ISM obscurer will maintain the AGN within the blow-out region, increasing the incidence of AGN in that region.
One could then consider the definition of a {\it forbidden} region with a redshift-dependent $N_{\rm H}$-threshold as displayed in Figure\,\ref{fig:sketch_forbidden_evolution}. With increasing redshift, the $N_{\rm H}$-threshold would increase and, subsequently, the intersection with the effective Eddington limit would be shifted towards higher\,$\lambda$ (horizontal large {\it red arrow} in Figure\,\ref{fig:sketch_forbidden_evolution}). This intersection corresponds to the $\lambda_{\rm break}$ at which the obscured AGN fraction strongly decreases. 
Indeed, only AGN with higher accretion rates $\lambda>\lambda_{\rm break}$ and higher obscuration $N_{\rm H}>N_{\rm H,\,ISM}$ can potentially be impacted by the dusty radiation pressure and have their obscurer rapidly blown away, decreasing the observed obscured AGN fraction.
It is important to note, however, that the $\lambda_{\rm break}$ values shown in Figure\,\ref{fig:sketch_forbidden_evolution} should only be considered illustrative. As mentioned earlier, the effective Eddington limit curve is not universal. Moreover, the $N_{\rm H,\, ISM}$ redshift evolution used here \citep{Gilli_2022} corresponds to the median ISM obscuration, i.e. more heavily ISM-obscured AGN can still occupy the redefined redshift-dependent {\it forbidden} region. 
Nevertheless, this scenario explains qualitatively the increasing $f_{\rm blow-out}$ with redshift (Figure\,\ref{fig:f_blowout_z_evolution}) and the shift of $\lambda_{\rm break}$ toward higher $\lambda$ with redshift (Figure\,\ref{fig:f_obs_z_evolution}).

\section{Summary and future works}\label{Sec:Summary}

In this study, we investigate the incidence of obscuration on the growth of AGN and its evolution up to redshift $z\leq 3$ by combining data from three X-ray field surveys: COSMOS, AEGIS, CDFS (Section\,\ref{Sec:Data_reduc}). The X-ray obscuration and intrinsic luminosity are measured by X-ray spectroscopy, while the stellar mass of the host galaxy is estimated through SED fitting (Section\,\ref{subsec:xray_spectroscopy} and \ref{subsec:SED_fitting}, respectively). 
These measurements are combined through a non-parametric Bayesian analysis, with careful propagation of measurement uncertainties throughout the analysis, to constrain the SARD (Section\,\ref{subsec:sBHAR_calculation}). 
Pioneer studies [\citetalias{Ricci_2017, Ricci_2022, Ananna_2022b}], using hard-X-ray ($>10$ keV) selected AGN, have investigated the relation between obscuration and accretion rate. However, our analysis expands this investigation, for the first time, to higher redshift $\sim 3$ and with a larger AGN sample ($\sim 4000$ sources). Furthermore, an innovative aspect of our study is its non-parametric approach, eliminating dependencies on prior assumptions regarding the shape and evolution of the SARD.

The key results of our research are summarised below
\begin{itemize}
    \item Unobscured and obscured SARDs have similar shapes at first approximation, in agreement with the AGN orientation model paradigm (Figure\,\ref{fig:sBHAR_z_evol}).
    
    \item The SARDs of both unobscured and obscured AGN populations shift toward higher accretion rates with redshift (Figure\,\ref{fig:sBHAR_z_evol}) driving a systematic increase in the mean accretion rate (Figure\,\ref{fig:mean_lambda}). These results are consistent with the AGN "downsizing" and suggest that the driver of these increases similarly impacts both AGN populations, possibly due to the increased gas availability in galaxies at higher redshifts (see discussion in Section\,\ref{discussion_redshift}).
    
    \item When normalised over $N_{\rm H}$ (Figure\,\ref{fig:sBHAR_normedNH}), the SARD of unobscured AGN show a small systematic offset towards higher accretion rates compared to the obscured population, leading to a $\sim 0.5$\,dex offset in <log$\lambda$> (Figure\,\ref{fig:mean_lambda}). These findings argue against the simplest unification model.
    
    \item For all redshifts, the obscured AGN fraction is high before declining at a certain $\lambda_{\rm break}$ value. 
    Below $z<0.5$, our findings behave similarly to the results of \citetalias{Ricci_2017} up to log$\lambda<-1$. 
    However, the $\lambda_{\rm break}$ shifts towards higher\,$\lambda$ with redshift, leading to an increased obscured AGN fraction at high-$\lambda$ with redshift.
    
    \item In the $\lambda-N_{\rm H}$ plane's {\it forbidden} region, where radiation pressure is thought to efficiently clear obscuration, a sparse population is expected (\citetalias{Ricci_2017}), as indicated in our observations at $z<0.5$. However, our results demonstrate a gradual increase in AGN density within this region at higher redshifts (see Figure\,\ref{fig:blowout_region_evol} and\,\ref{fig:f_blowout_z_evolution}). 

    \item The redshift evolution of the obscured AGN fraction and the {\it forbidden} region AGN fraction may be attributed to either an increased incidence of outflows in AGN (See discussion in Section\,\ref{discussion_orientation}) or to an increased host-galaxy interstellar medium obscuration with redshift (see discussion in Section\,\ref{discussion_ISM} and Figure\,\ref{fig:sketch_forbidden_evolution}).
\end{itemize}

In conclusion, our novel approach, differentiating between obscured and unobscured AGN populations, provides valuable insights into the relationship between obscuration and SMBH accretion. The similarity of the overall SARD shape and the evolution of the mean accretion rate between obscured and unobscured AGN suggest a shared underlying physical driver, broadly supporting the orientation model. However, at higher accretion rates, a stark divergence emerges, indicating a more complex behaviour. The evolution of this divergence as a function of the redshift can shed light on the growth and outflow mechanisms that drive the AGN life cycle.

Follow-up studies could assess the high outflow incidence expected in {\it forbidden} region AGN as a function of the redshift to inspect this transitioning stage of AGN. For that purpose, several wavebands can be valuable, for instance, ALMA can also trace molecular outflows. Additionally, optical spectra can inspect the presence of broad absorption lines \citep[BAL, ][]{Weymann_1991, Gallagher_2006}, characteristic of outflows. 
An obvious limitation would be the availability of optical spectra, especially since obscuration strongly impacts the emission lines, but future large spectroscopic surveys like MOONRISE \citep{Maiolino_2020}, the 4MOST AGN survey \citep[][ Andonie in prep.]{Merloni_2019} or with the MOSAIC instrument on ELT will provide statistically significant spectra samples.
Moreover, the recent work of \cite{Fawcett_2023} suggests that outflows interacting with the ISM could produce radio-quiet emission, linking radio-detection fraction and obscuration. A follow-up analysis of the radio-detection in the forbidden region is in preparation.
Complementary, following the methodology of \cite{Circosta_2019, Andonie_2023}, ALMA observations of blow-out region AGN could provide constraints on the ISM-obscuration, and doing so, remove ISM-obscured AGN and refine the {\it forbidden} region selection.
Such a rigorous and systematic multi-wavelength analysis of these AGN will provide valuable knowledge on the physics linking obscuration and accretion across cosmic time.

\section*{Acknowledgements}

The research leading to these results has received funding from the EU H2020-MSCA-ITN-2019 Project 860744 "BiD4BESt: Big Data applications for black hole Evolution STudies" and the Hellenic Foundation for Research and Innovation (HFRI) project "4MOVE-U" grant agreement 2688, which is part of the programme "2nd Call for HFRI Research Projects to support Faculty Members and Researchers".

This research made use of Astropy,\footnote{http://www.astropy.org} a community-developed core Python package for Astronomy \citep{astropy:2013, astropy:2018}. 
This research has made use of data obtained from the Chandra Data Archive and the Chandra Source Catalog, and software provided by the Chandra X-ray Center (CXC) in the application packages CIAO and Sherpa.
For analysing X-ray spectra, we use the analysis software BXA \citep{Buchner_2014}, which connects the nested sampling algorithm UltraNest \citep{Buchner_2021} with the fitting environment CIAO/Sherpa \citep{Fruscione_2006}.

DMA additionally thanks the Science Technology Facilities Council (STFC) for support from the Durham consolidated grant (ST/T000244/1).
AL was partially funded by "Data Science methods for MultiMessenger Astrophysics \& Multi-Survey Cosmology" funded by the Italian Ministry of University and Research, Programmazione triennale 2021/2023 (DM n.2503 dd. 9 December 2019), Programma Congiunto Scuole.
RCH acknowledges support from NASA through ADAP grant number 80NSSC23K0485.
JA acknowledges support from UKRI grant MR/T020989/1.

For the purpose of open access, the authors have applied a Creative Commons Attribution (CC BY) licence to any Author Accepted Manuscript version arising from this submission.

\section*{Data Availability}

We release the Stan products of the SARD calculations along with the scripts to exploit and display the data. The files are described and available at \href{https://doi.org/10.5281/zenodo.10797147}{https://doi.org/10.5281/zenodo.10797147}.

All other data products used in this paper are available on-demand to the authors.



\bibliographystyle{mnras}
\bibliography{Main} 




\appendix

\section{SED fitting module parameters}
In this appendix, we summarize in Table\,\ref{tab:cigale_params} the different modules used for the SED fitting in Section\,\ref{subsec:SED_fitting}. The range of the different parameters is also indicated.

\begin{table*}
    \centering
    \caption{The modules and their parameters along their grid values used by CIGALE for the SED fitting of our sources (Section\,\ref{subsec:SED_fitting}).}
    \begin{tabular}{c|c|c}
        \hline\hline
         Parameter & Module & Value  \\
         \hline\hline
          & Star formation history & \\
            & Delayed SFH with recent burst & \\
          Age of the main population &  & 500, 1000, 3000, 5000, 7000, 9000 Myr\\
           e-folding time main population &  & 500, 1500, 3000, 5000, 7000 Myr\\
            Age of the burst &  & 20, 200 Myr\\
             e-folding time of the burst &  & 50 Myr\\
              Burst stellar mass fraction &  & 0.0, 0.1\\
        \hline \hline
          & Simple Stellar population & \\
           & \protect\cite{Bruzual_2003} & \\ 
           Initial Mass Function &  & \protect\cite{Chabrier_2003} \\
            Metallicity &  & 0.02 \\
        \hline \hline
         & Galactic dust extinction & \\
          & \protect\cite{Calzetti_2000} & \\
          E(B-V)$_{young}$ &  & 0.0, 0.1, 0.25, 0.5, 0.75, 0.9\\
           UV$_{bump}^{\lambda}$ &  & 217.5 nm \\
        \hline \hline
        & Galactic dust emission & \\
        & \protect\cite{Dale_2014} & \\
        $\alpha$ slope in $dM_{dust} \propto U^{-\alpha}dU$&  & 1.0, 1.5, 2.0, 2.5, 3\\
        \hline \hline
        & Nebular & \\
        & \protect\cite{Inoue_2011} & \\
        log U &  & -2.0 \\
        width lines&  & 200 km.s$^{-1}$ \\
        \hline \hline
        & AGN module: SKIRTOR &   \\
        & \protect\cite{Stalevski_2016} & \\
        Torus optical depth at 9.7 microns &  & 3.0, 7.0\\
        Torus density radial parameter &  & 1.0\\
        Torus density angular parameter&  & 1.0 \\
        Angle between the equatorial plan and edge of the torus &  & 40$^{\circ}$ \\
        Ratio of the maximum to minimum radii of the torus &  & 20\\
        Viewing angle &  & 30$^{\circ}$, 70$^{\circ}$\\
        Disk spectrum &  & \protect\cite{Schartmann_2005} \\
        Power-law index modifying the optical slope of the disk &  & -0.36\\
        AGN fraction &  & 0.01, 0.1, 0.2, 0.3, 0.4, 0.5, 0.6, 0.7, 0.8, 0.9, 0.99\\
        Extinction law of polar dust &  & SMC\\
        E(B - V) of polar dust &  & 0.0, 0.02, 0.1, 0.5, 1.0\\
        Temperature of polar dust &  & 100 K \\
        Emissivity of polar dust &  & 1.6 \\
        \hline \hline
        Total number of models & & 792 000\\
        \hline \hline
        
    \end{tabular}
    \label{tab:cigale_params}
\end{table*}

\section{Physical parameter distribution}\label{appendix_B}

With the X-ray spectroscopy and SED fitting methodologies described in Sections\,\ref{subsec:xray_spectroscopy} and \ref{subsec:SED_fitting}, for the whole AGN sample, we recover the intrinsic X-ray luminosity $L_{\rm X}$, column density $N_{\rm H}$ and stellar mass $M_\star$, from which we can compute $\lambda \propto L_{\rm X}/M_\star$.

\begin{figure}
    \centering
    \includegraphics[width=0.7\linewidth]{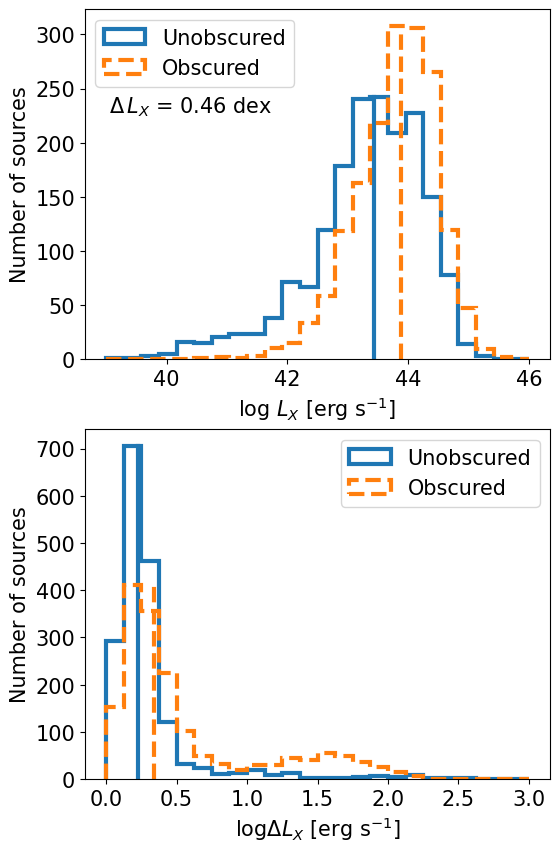}
    \caption{{\it Top panel}: Logarithmic intrinsic 2-10\,keV X-ray luminosity distribution for obscured ({\it orange dashed}) and unobscured ({\it solid blue}) AGN. The median value for both samples is also indicated by a vertical line of the corresponding colour and style.
    {\it Bottom panel}: Distribution of intrinsic X-ray luminosity uncertainties for unobscured and unobscured AGN. The colour code is identical to in the top panel.}
    \label{fig_app:Lx_hist}
\end{figure}

Figure\,\ref{fig_app:Lx_hist} displays the distribution of intrinsic 2-10\,keV X-ray luminosities for both obscured and unobscured AGN. As expected, obscured AGN have a higher median luminosity, by $\sim 0.5$ dex. This is due to a selection effect, for a given flux limit at the same redshift, an obscured AGN must be brighter than its unobscured counterpart to be detected. The bottom panel of figure\,\ref{fig_app:Lx_hist} shows the 1-$\sigma$ uncertainty distribution of the X-ray luminosity measurements. One can notice a bump in the obscured distribution, around 1.5. This is due to the double-peaked sources mentioned in \cite{Laloux_2023}. For these sources, the $L_{\rm X}$ and $N_{\rm H}$ posteriors show two distinct peaks, one in the CTN obscured regime and one in the CTK regime, which corresponds, respectively, to the moderate and high luminosity peaks. As a consequence, the 1-$\sigma$ uncertainty of the luminosity is broad, which explains the $\log\Delta L_{\rm X}$ bump around 1.5 dex.

\begin{figure}
    \centering
    \includegraphics[width=0.7\linewidth]{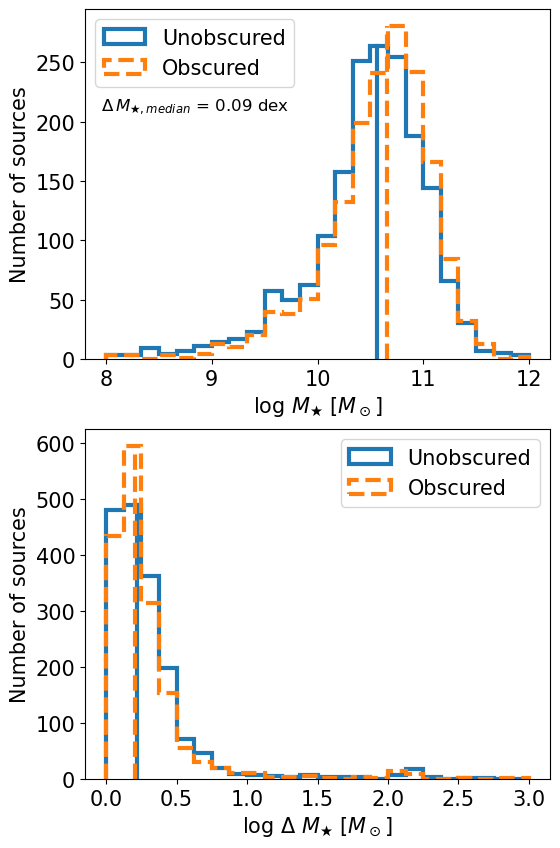}
    \caption{{\it Top panel}: Logarithmic stellar mass distribution for obscured ({\it orange dashed}) and unobscured ({\it solid blue}) AGN. The median value for both samples is also indicated by a vertical line of the corresponding colour and style.
    {\it Bottom panel}: Distribution of stellar mass uncertainties for unobscured and unobscured AGN. The colour code is identical to in the top panel.
    }
    \label{fig_app:Mstar_hist}
\end{figure}

In the same way, Figure\,\ref{fig_app:Mstar_hist} displays the distribution of log$M_\star$ measurements and uncertainties for both AGN populations. The median log$M_\star$ of obscured AGN is moderately higher than for unobscured AGN by $\sim 0.1$dex. This could be explain by the fact that more luminous AGN are slightly more massive than their less massive counterparts, and as seen in Figure\,\ref{fig_app:Lx_hist}, obscured AGN are in average more luminous than unobscured AGN. Nevertheless, the stellar mass uncertainty of both populations is very similar, showing no systematic offset between one or the other population. This result shows that there is no specific bias in terms of host galaxy stellar mass constraints between obscured and unobscured AGN.

\begin{figure}
    \centering
    \includegraphics[width=0.7\linewidth]{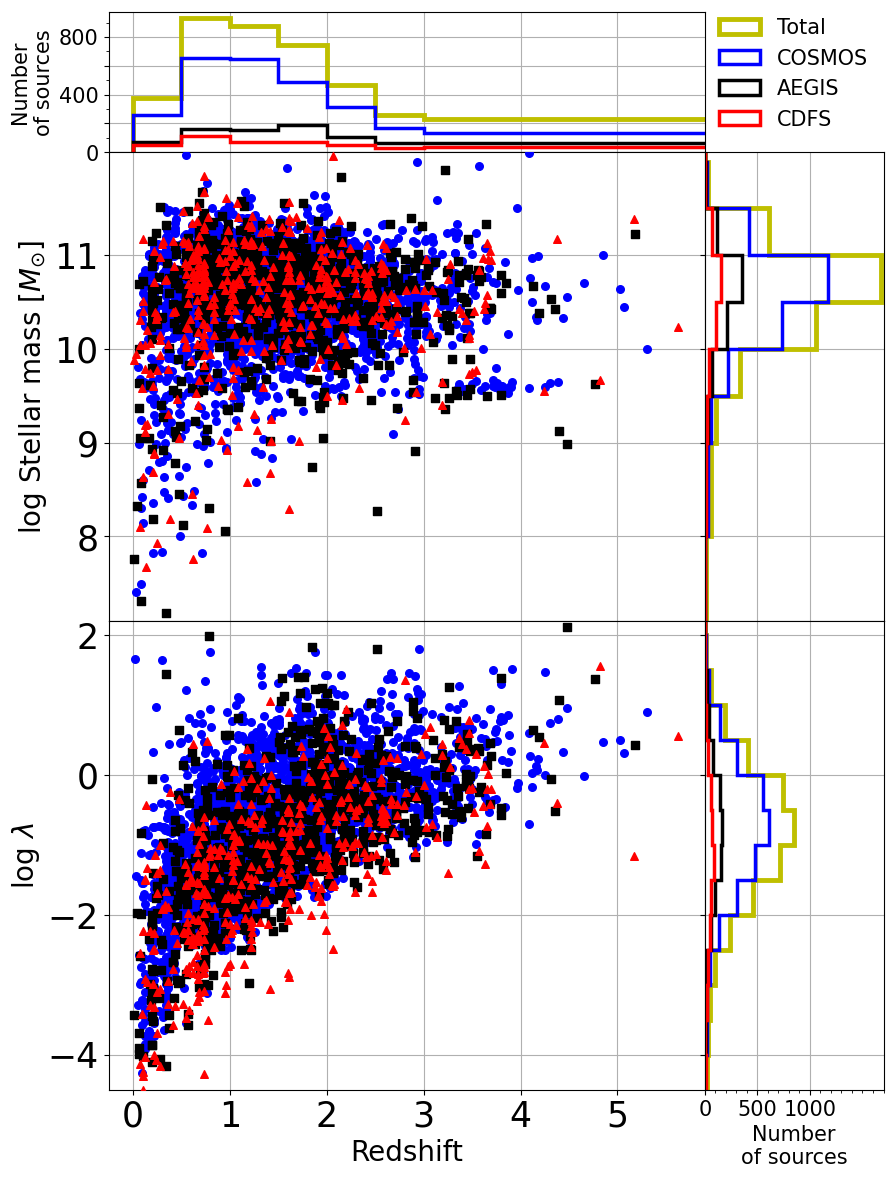}
    \caption{{\it Top panel}: Redshift distribution of the different surveys with COSMOS ({\it blue}), AEGIS ({\it black}), CDFS ({\it red}) and the total sample ({\it yellow}).
    {\it Central left panel}: Distribution of our sample stellar mass as a function of the redshift. The blue circles, the black squares and the red triangles correspond to the AGN of the COSMOS Legacy, AEGIS and CDFS surveys, respectively.
    {\it Central right panel}: Stellar mass distribution of the different surveys. Same colour code as on the top panel.
    {\it Bottom left panel}: Distribution of our sample accretion rate $\lambda\propto L_{\rm X}/M_\star$ as a function of the redshift. The marker style and colour are identical to the central left panel.
    {\it Bottom right panel}: log$\lambda$ distribution of the different surveys. Same colour code as on the top panel.
    }
    \label{fig_app:Mstar_lambda_z_distrib}
\end{figure}

Finally, Figure\,\ref{fig_app:Mstar_lambda_z_distrib} displays the AGN distribution in the $M_\star-z$ and $\lambda-z$ plane for the three different surveys (COSMOS, AEGIS and CDFS). For clarity, in these plots, we do not show uncertainties, only the posterior median values.

\section{Duty cycle}\label{sec:appendix-DC}
In this appendix section, we present the duty cycle estimated in Section\,\ref{subsec:blowout_region}. It represents the probability that a galaxy hosts an AGN accreting faster than a specific accretion rate. It is computed as the integration of the SARD along the $\lambda$-axis with an increasing lower limit.

\begin{figure}
    \centering
    \includegraphics[width=1\linewidth]{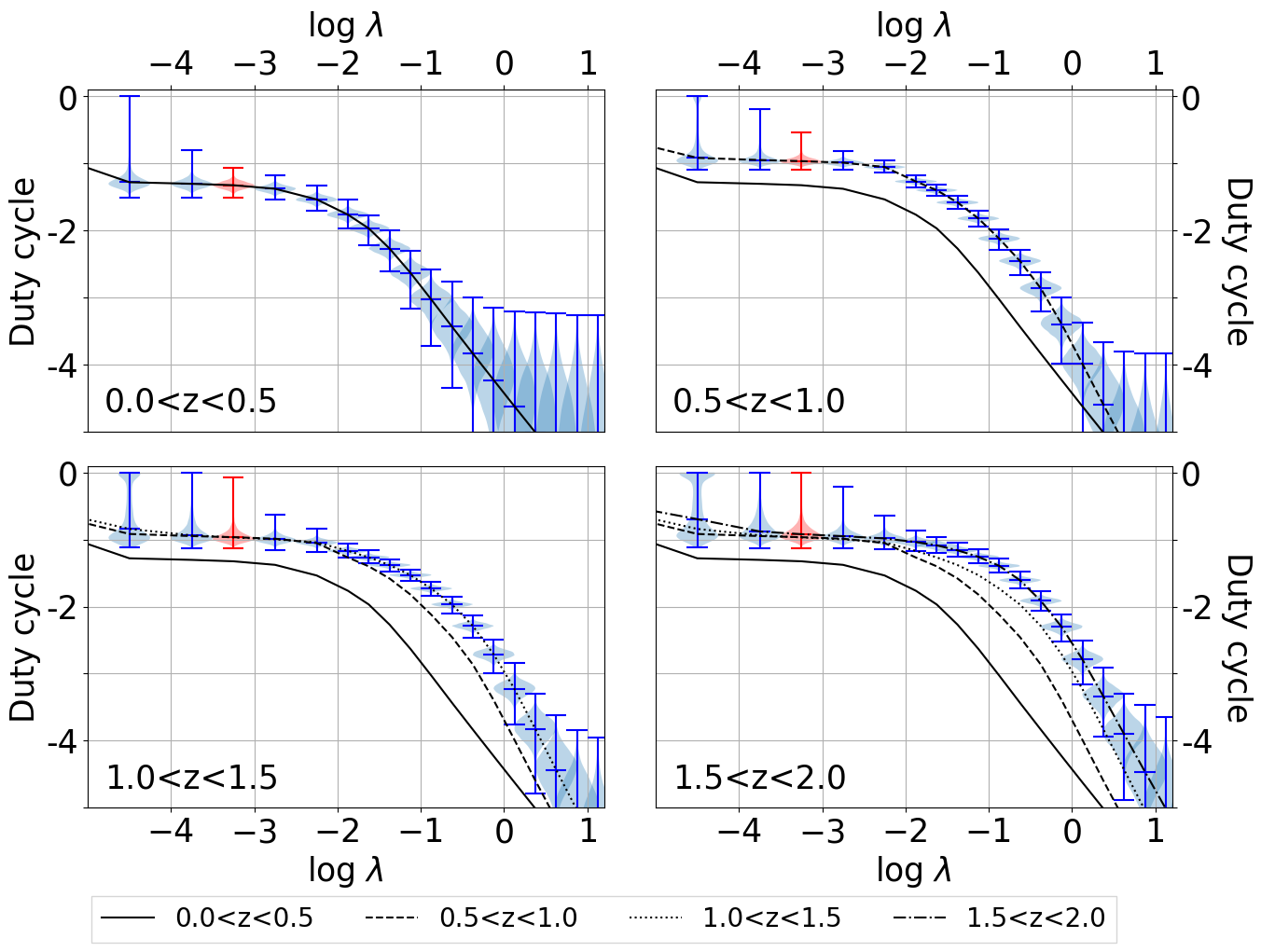}
    \caption{Duty cycle for different redshift intervals. It corresponds to the probability that a galaxy hosts an AGN accreting faster than a specific accretion rate. The {\it violin plots} represent the duty cycle posterior distribution. The different {\it black lines} represent the median duty cycle of the corresponding redshift interval.
    The {\it red} duty cycle is $P({\rm log}\lambda>-3.5)$ used in Section\,\ref{subsec:blowout_region} to compute $f_{\rm blow-out}$.}
    \label{fig_app:duty_cycle}
\end{figure}


\bsp	
\label{lastpage}
\end{document}